\documentclass[11pt,a4paper]{article}
\pdfoutput=1

\usepackage{
	jcappub,
	mathtools,
	microtype,
	upgreek,
	xcolor,
	xparse
}

\usepackage[caption=false]{subfig}

\NewDocumentCommand \secref{m}{section \ref{#1}}
\NewDocumentCommand \appref{m}{appendix \ref{#1}}
\NewDocumentCommand \figref{m}{figure \ref{#1}}

\let\usualpi\pi

\NewDocumentCommand \ii{}{\mathrm{i}}
\NewDocumentCommand \e{}{\mathrm{e}}
\RenewDocumentCommand \d{}{\mathrm{d}}
\NewDocumentCommand \I{}{\mathrm{I}}
\RenewDocumentCommand\pi{}{\uppi}

\NewDocumentCommand \m{}{\mathrm{m}}
\NewDocumentCommand \R{}{\textsc{r}}
\NewDocumentCommand \A{}{\textsc{a}}

\NewDocumentCommand \C{}{\mathrm{c}}

\RenewDocumentCommand \H{}{\mathcal{H}}
\NewDocumentCommand \J{}{\mathcal{J}}

\RenewDocumentCommand \L{}{\mathcal{L}}

\NewDocumentCommand \F{}{\mathcal{F}}

\RenewDocumentCommand \theta{}{\vartheta}

\NewDocumentCommand \D{}{\mathcal{D}}

\NewDocumentCommand \vect{m}{\vec{#1}}
\NewDocumentCommand \tens{m}{\boldsymbol{#1}}

\DeclareMathOperator{\tr}{\mathrm{tr}}

\DeclareMathOperator{\dirac}{\updelta_\textsc{d}}

\DeclareMathOperator{\kronecker}{\updelta}

\DeclareMathOperator{\step}{\uptheta}

\NewDocumentCommand \id{}{\mathcal{I}}

\NewDocumentCommand \upi{m}{\int \D #1 \,}

\NewDocumentCommand \bpi{mmm}{\int\limits_{#2}^{#3} \D #1 \,}

\NewDocumentCommand \usi{m}{\int \d #1 \,}

\NewDocumentCommand \bsi{mmm}{\int\limits_{#2}^{#3} \d #1 \,}

\NewDocumentCommand \bsiexp{mmm}{\int_{#2}^{#3} \d #1}

\NewDocumentCommand \umi{mm}{\int \d^{#1} #2 \,}

\NewDocumentCommand \fmi{mm}{\int \frac{\d^{#1} #2}{(2\pi)^{#1}} \,}

\NewDocumentCommand \is {}{\hspace{-1ex}}

\NewDocumentCommand \der{O{}m}{\frac{\updelta{#1}}{\ii \updelta{#2}}}

\NewDocumentCommand \derarg{O{}m}{\tfrac{\updelta #1}{\ii \updelta #2}}

\NewDocumentCommand \derao{O{}mm}{\frac{\updelta^{#3}#1}{\ii \updelta#2^{#3}}}

\NewDocumentCommand \deraoarg{O{}mm}{\tfrac{\updelta^{#3}#1}{\ii \updelta#2^{#3}}}

\NewDocumentCommand \td{mm}{\frac{\d #1}{\d #2}}

\NewDocumentCommand \tdarg{O{}m}{\tfrac{\d #1}{\d #2}}

\NewDocumentCommand \pdao{O{}mm}{\frac{\partial^{#3}#1}{\partial#2^{#3}}}

\NewDocumentCommand \pdarg{O{}m}{\tfrac{\partial #1}{\partial #2}}

\NewDocumentCommand \fd{mm}{\frac{\updelta #1}{\updelta #2}}

\NewDocumentCommand \fdao{O{}mm}{\frac{\updelta^{#3}#1}{\updelta#2^{#3}}}

\NewDocumentCommand \fdaoarg{O{}mm}{\tfrac{\updelta^{#3}#1}{\updelta#2^{#3}}}

\RenewDocumentCommand \={}{\hphantom{=}\hspace{0.5ex}}
\NewDocumentCommand \+{}{\hphantom{+}}
\NewDocumentCommand \half{}{\hphantom{\tfrac{1}{2}}}

\NewDocumentCommand \contr{}{\cdot}

\NewDocumentCommand \IC{}{\usi{\Gamma_\ii}}

\NewDocumentCommand \ini{}{{(\ii)}}

\NewDocumentCommand \free{}{{(0)}}

\NewDocumentCommand \cl{}{{(\mathrm{cl})}}

\NewDocumentCommand \largescale{}{{(\text{ls})}}

\NewDocumentCommand \prop{}{{(\Prop)}}

\NewDocumentCommand \proplargescale{}{{(\Prop, \text{ls})}}

\NewDocumentCommand \Zcoll{}{Z_{\Phi}}
\NewDocumentCommand \fZcoll{}{Z_{\Phi,0}}
\NewDocumentCommand \fdZcoll{}{Z_{\tilde{\Phi},0}}
\NewDocumentCommand \Wcoll{}{W_{\Phi}}
\NewDocumentCommand \fWcoll{}{W_{\Phi,0}}
\NewDocumentCommand \fdWcoll{}{W_{\tilde{\Phi},0}}
\NewDocumentCommand \Zmac{}{Z_{\phi}}

\NewDocumentCommand \Wmac{}{W_{\phi}}

\NewDocumentCommand \G{m}{G_{#1}}
\NewDocumentCommand \fG{}{G^\free}

\NewDocumentCommand \Grr{}{G_{\rho \rho}}
\NewDocumentCommand \Gff{}{G_{f f}}

\NewDocumentCommand \fGf{}{G^\free_f}

\NewDocumentCommand \fGff{}{G^\free_{f f}}

\NewDocumentCommand \fGfB{}{G^\free_{f B}}
\NewDocumentCommand \fGfF{}{G^\free_{f \mathcal{F}}}

\NewDocumentCommand \fGBf{}{G^\free_{B f}}
\NewDocumentCommand \fGFf{}{G^\free_{\mathcal{F} f}}

\NewDocumentCommand \fGFF{}{G^\free_{\mathcal{F} \mathcal{F}}}
\NewDocumentCommand \afGff{}{\tilde{G}^\free_{ff}}
\NewDocumentCommand \afGfF{}{\tilde{G}^\free_{f \mathcal{F}}}

\NewDocumentCommand \cosmoT{}{T}

\NewDocumentCommand \sigfB{}{\sigma_{f B}}
\NewDocumentCommand \sigBf{}{\sigma_{B f}}

\NewDocumentCommand \Prop{}{\Delta}
\NewDocumentCommand \Propff{}{\Delta_{f f}}
\NewDocumentCommand \Propfb{}{\Delta_{f \beta}}
\NewDocumentCommand \Propbf{}{\Delta_{\beta f}}
\NewDocumentCommand \Propbb{}{\Delta_{\beta \beta}}

\NewDocumentCommand \iProp{}{\Delta^{-1}}
\NewDocumentCommand \iPropff{}{(\Delta^{-1})_{f f}}
\NewDocumentCommand \iPropfb{}{(\Delta^{-1})_{f \beta}}
\NewDocumentCommand \iPropbf{}{(\Delta^{-1})_{\beta f}}
\NewDocumentCommand \iPropbb{}{(\Delta^{-1})_{\beta \beta}}

\NewDocumentCommand \PropR{}{\Delta_{\mathrm{R}}}
\NewDocumentCommand \PropA{}{\Delta_{\mathrm{A}}}

\NewDocumentCommand \aPropR{}{\tilde{\Delta}_{\mathrm{R}}}
\NewDocumentCommand \aPropA{}{\tilde{\Delta}_{\mathrm{A}}}

\RenewDocumentCommand \Vert{}{\mathcal{V}}

\begin{document}

\title{Resummed Kinetic Field Theory: general formalism and linear structure growth from Newtonian particle dynamics}

\author[a,1]{Robert Lilow,\note{Present address: Department of Physics, Technion, Haifa 3200003, Israel}}
\author[a]{Felix Fabis,}
\author[a]{Elena Kozlikin,}
\author[a,b]{Celia Viermann}
\author[a]{and Matthias Bartelmann}

\affiliation[a]{Heidelberg University, Zentrum f\"ur Astronomie, Institut f\"ur Theoretische Astrophysik, Philosophenweg 12, 69120 Heidelberg, Germany}
\affiliation[b]{Heidelberg University, Kirchhoff-Institut f\"ur Physik, Im Neuenheimer Feld 227, 69120 Heidelberg, Germany}

\emailAdd{rlilow@campus.technion.ac.il}
\emailAdd{felix.fabis@posteo.de}
\emailAdd{elena.kozlikin@uni-heidelberg.de}
\emailAdd{celia.viermann@kip.uni-heidelberg.de}
\emailAdd{bartelmann@uni-heidelberg.de}

\abstract{In earlier work, we have developed a nonequilibrium statistical field theory description of cosmic structure formation, dubbed Kinetic Field Theory (KFT), which is based on the Hamiltonian phase-space dynamics of classical particles and thus remains valid beyond shell-crossing. Here, we present an exact reformulation of the KFT framework that allows to resum an infinite subset of	terms appearing in the original perturbative expansion of KFT. We develop the general formalism of this resummed KFT, including a diagrammatic language for the resummed perturbation theory, and compute the lowest-order results for the power spectra of the dark matter density contrast and momentum density. This allows us to derive analytically how the linear growth of the largest structures emerges from Newtonian particle dynamics alone, which, to our knowledge, is the first time this has been achieved.}


\arxivnumber{1809.06942}

\maketitle

\flushbottom

\section{Introduction}
\label{sec:intro}
The statistical properties of cosmic large-scale structures can be used to study the nature of gravity and test our cosmological model. This, however, requires a precise and accurate theoretical understanding of the evolution of self-gravitating systems, which has proven to be an extremely challenging task.

While conventional Eulerian and Lagrangian analytic descriptions of structure formation provide precise predictions on linear and mildly nonlinear scales, it has been notoriously difficult for them to reach further into the nonlinear regime \cite{bernardeau_large-scale_2002}. This is largely due to the fact that they assume matter to move along a unique velocity field. With collisionless dark matter dominating cosmic structures, though, particle trajectories can cross and form multiple streams, at which point the assumption of uniqueness breaks down. Numerical $N$-body simulations, on the other hand, are able to describe the nonlinear small-scale dynamics, as they follow the phase-space trajectories of individual tracer particles. But on the downside, they offer only little insight into the underlying principles of structure formation and are computationally expensive.

To overcome these problems, we have developed a new analytic description of cosmic structure formation, dubbed Kinetic Field Theory (KFT), which is built on the full Hamiltonian dynamics of classical particles in phase-space and can thus be seen as an analytic analogue to an $N$-body simulation. At its very core, KFT is based on the Martin-Siggia-Rose formalism for describing classical systems in terms of a path integral approach \cite{martin_statistical_1973}. This was first applied to Hamiltonian dynamics by Gozzi et al.~in \cite{gozzi_hidden_1988,gozzi_hidden_1989} and further developed into a full field-theoretical description of kinetic theory by Mazenko and Das in \cite{mazenko_fundamental_2010,das_field_2012}. Building on these pioneering works, we extended the KFT framework to describe the dynamics of an initially correlated ensemble of particles and applied it to cosmic structure formation in \cite{bartelmann_microscopic_2016,bartelmann_kinetic_2017,fabis_kinetic_2018}.

An unusual aspect of KFT, when compared to other nonequilibrium statistical field theories, is that the fundamental dynamical fields in KFT are of microscopic nature even though we are actually interested in macroscopic fields like the density contrast or the momentum density. This renders the application of many well-established and powerful field-theoretical tools, e.\,g. renormalization group techniques \cite{berges_non-perturbative_2002,matarrese_resumming_2007,floerchinger_renormalization-group_2017}, rather difficult. To facilitate the use of such tools in the future, we develop here an exact reformulation of the KFT path integral in terms of macroscopic fields while preserving all information on the microscopic dynamics. 

We show how this reformulation also gives rise to a new macroscopic perturbation theory that resums an infinite subset of contributions to the original microscopic perturbative expansion in orders of the interaction potential presented in \cite{bartelmann_microscopic_2016}. To systematise the calculation of macroscopic-field cumulants within this resummed KFT (RKFT) framework, we introduce a Feynman diagram representation similar to the one used in \cite{crocce_renormalized_2006}.

Using RKFT enables us to treat dark matter particles in terms of their fundamental Newtonian dynamics rather than the modified version of Zel'dovich dynamics we have been using so far \cite{bartelmann_trajectories_2015}. In previous works, we exploited that the inertial motion of particles on Zel'dovich-type trajectories already contains part of the gravitational interaction. Due to this, even a first-order calculation in the microscopic perturbation theory reproduces the nonlinear evolution of the density contrast power spectrum known from numerical simulations over a wide range of scales remarkably well \cite{bartelmann_microscopic_2016}. For Newtonian dynamics this is not the case and any expansion to finite order in the Newtonian gravitational potential does not even reproduce the linear growth of structures correctly. Using RKFT allows to overcome this limitation. In particular, we show how the lowest order in the new macroscopic perturbation theory precisely recovers the linear large-scale growth by partially resumming the gravitational interactions between particles following Newtonian dynamics. To our knowledge, this is the first time this has been achieved.

In \secref{sec:micro}, we briefly review the general framework of KFT by setting up the path integral for classical particles and explaining how cumulants of macroscopic fields can be obtained from it. We also summarise how these cumulants can be calculated using the microscopic perturbative expansion in orders of the interaction potential. Afterwards, we show in \secref{sec:macro} how the KFT path integral can be reformulated in terms of macroscopic fields, yielding the new RKFT framework. Some physical intuition for the resummation process at the level of the microscopic particle dynamics is given and the properties of the new macroscopic perturbation theory are discussed in this section. For this purpose, we introduce a diagrammatic language and derive the respective Feynman rules. In \secref{sec:cosmo}, the RKFT framework is used to describe dark matter structure formation in a standard $\Lambda$CDM cosmology. We specifically compute the lowest-order results for the power spectra of the density contrast and momentum density, demonstrating how the linear growth of the largest structures emerges from Newtonian particle dynamics. Finally, we summarise our results and conclusions in \secref{sec:conclusions}. Some of the more technical aspects of our derivations can be found in the appendices.

\section{General framework of KFT}
\label{sec:micro}
To render this paper as self-contained as possible, we give a brief review of the general framework of KFT here, focussing on the key aspects necessary to understand the subsequent sections. For a more detailed description we refer to \cite{bartelmann_microscopic_2016,fabis_kinetic_2018,das_field_2012}.

\subsection{Path integral for classical particles}
\label{sec:micro:path_integral_for_classical_N-particle_system}
Consider a classical $N$-particle system described by the phase-space trajectories $\vect{x}_j(t)=\bigl(\vect{q}_j(t), \vect{p}_j(t)\bigr)$ of its individual particles, ${j=1,\dotsc,N}$. Here $\vect{q}_j(t)$ and $\vect{p}_j(t)$ are the position and momentum of the $j$-th particle, respectively. To condense the notation, we will combine these single-particle vectors into $N$-particle tensors $\tens{x}(t)$, $\tens{q}(t)$ and $\tens{p}(t)$, adopting the conventions introduced in \cite{bartelmann_microscopic_2016} and \cite{fabis_kinetic_2018},
\begin{equation}
	\tens{a}(t) \coloneqq \sum_{j=1}^N \vect{a}_j(t) \otimes \vect{e}_j \,, \qquad \tens{a \contr b} \coloneqq \sum_{j=1}^N \, \bsi{t}{t_\ii}{\infty} \, \vect{a}_j(t) \cdot \vect{b}_j(t) \,,
\end{equation}
where $\vect{e}_j$ is the $N$-dimensional Cartesian base vector with entries $\bigl(\vect{e}_j\bigr)_i = \kronecker_{ij}$ and $t_\ii$ denotes the initial time.

The initial phase-space configuration $\tens{x^\ini} \coloneqq \tens{x}(t_\ii)$ of such an $N$-particle system is assumed to be characterised by a probability distribution $\mathcal{P}_\ii\bigl(\tens{x^\ini}\bigr)$ and its dynamics at times $t \geq t_\ii$ shall be governed by some equations of motion $\tens{E}[\tens{x}(t)] = 0$. A central idea of KFT is to encapsulate both of these in the partition function
\begin{equation}
	Z = \usi{\tens{x^\ini}} \mathcal{P}_\ii\bigl(\tens{x^\ini}\bigr) \bpi{\tens{x}}{\tens{x^\ini}}{} \dirac\bigl[\tens{x}(t) - \tens{x^\cl}\bigl(\tens{x^\ini},t\bigr)\bigr] \,.
	\label{eq:micro:definition_of_microscopic_path_integral_in_terms_of_EOM_in_terms_of_classical_trajectory}
\end{equation}
Here, the dynamics is incorporated by functionally integrating over all possible phase-space trajectories $\tens{x}(t)$ starting from an initial configuration $\tens{x}^\ini$, where a functional Dirac delta distribution ensures that only the classical $N$-particle trajectory $\tens{x^\cl}\bigl(\tens{x^\ini},t\bigr)$, which solves the equations of motion, contributes. The stochastic initial conditions are then taken into account by averaging over the initial probability distribution $\mathcal{P}_\ii\bigl(\tens{x^\ini}\bigr)$.

Since the full solution of the equations of motion is generally not known, we re-express the delta distribution in \eqref{eq:micro:definition_of_microscopic_path_integral_in_terms_of_EOM_in_terms_of_classical_trajectory} in terms of the equations of motion themselves,
\begin{equation}
	\dirac\bigl[\tens{x}(t) - \tens{x^\cl}\bigl(\tens{x^\ini},t\bigr)\bigr] = \dirac\bigl[\tens{E}[\tens{x}(t)]\bigr] \, \left. \det\biggl[\fd{\tens{E}[\tens{x}(t)]}{\tens{x}(t')}\biggr] \, \right|_{\tens{x}(t) = \tens{x^\cl}(\tens{x^\ini},t)} \,.
	\label{eq:micro:transforming_delta_function_in_classical_trajectory_to_delta_function_in_EOM}
\end{equation}
In this work, we will only consider Hamiltonian dynamics for which the functional Jacobian determinant on the right-hand-side is just a constant that can be absorbed into the normalisation of the path integral measure \cite{das_field_2012,gozzi_hidden_1989}.\footnote{See \cite{mazenko_fundamental_2010} for an example of how to treat dynamics with non-constant Jacobian determinant.} The partition function can then be brought into the more convenient form
\begin{equation}
	Z	=	\IC \bpi{\tens{x}}{\tens{x^\ini}}{} \upi{\tens{\chi}} \, \e^{\ii \tens{\chi \cdot E}[\tens{x}]}
		\label{eq:micro:delta_trick_in_microscopic_path_integral}
\end{equation}
by expressing the delta distribution in terms of a functional Fourier integral with respect to an auxiliary field $\tens{\chi}(t)$ with components $\vect{\chi}_j(t) = \bigl(\vect{\chi}_{q_j}(t), \vect{\chi}_{p_j}(t)\bigr)$ conjugate to $\vect{x}_j(t)$ and defining the short-hand notation $\IC \coloneqq \usi{\tens{x^\ini}} \mathcal{P}_\ii\bigl(\tens{x^\ini}\bigr)$ for the initial phase-space average.

The equations of motion are given by Hamilton's equations,
\begin{equation}
	0 = \tens{E}[\tens{x}] = \dot{\tens{x}} - \tens{\J \, \nabla_x} \H[\tens{x}] \,, \qquad
	\tens{\J} \coloneqq
		\begin{pmatrix}
			0_3 & \id_3 \\
			-\id_3 & 0_3
		\end{pmatrix}
		\otimes \id_N \,,
	\label{eq:micro:Hamilton's_equations}
\end{equation}
where $\id_d$ denotes the $d$-dimensional identity matrix and $0_d$ a $d$-dimensional matrix with all entries being zero. The Hamiltonian shall take the form ${\H = \H_0 + \H_\I}$, with $\H_0$ describing the free evolution and $\H_\I$ the interactions. In addition, we assume on the one hand that the free equations of motion are linear in $\tens{x}$ and can hence be solved in terms of a retarded Green's function,
\begin{equation}
	\tens{x^\free}\bigl(\tens{x^\ini},t\bigr) \coloneqq \tens{x^\cl}\bigl(\tens{x^\ini},t\bigr) \Bigr|_{\tens{\H_\I}=0} = \tens{\mathcal{G}}(t,t_\ii) \, \tens{x^\ini} \,,
\label{eq:micro:free_trajectories_in_general_form}
\end{equation}
with
\begin{equation}
	\tens{\mathcal{G}}(t,t') \coloneqq
	\mathcal{G}(t,t') \otimes \id_N \,, \qquad
	\mathcal{G}(t,t') \coloneqq
	\begin{pmatrix}
		g_{qq}(t,t') \, \id_3 \;\;\; & g_{qp}(t,t') \, \id_3 \\
		g_{pq}(t,t') \, \id_3 \;\;\; & g_{pp}(t,t') \, \id_3
	\end{pmatrix}
	\propto \step(t - t') \,.
\label{eq:micro:retarded_Green's_function}
\end{equation}
On the other hand, we restrict ourselves to systems without external forces and assume that $\H_\I$ is the total potential energy generated by a superposition of single-particle potentials $v$ depending only on the spatial distances between particles and possibly on time,
\begin{equation}
	\H_\I[\tens{x}(t)] = \H_\I[\tens{q}(t)] \coloneqq \frac{1}{2} \sum_{j \neq k = 1}^N v\,\bigl(|\vect{q}_j-\vect{q}_k|, t\bigr) \,.
	\label{eq:micro:total_potential_energy}
\end{equation}

Let us further introduce the combined microscopic field ${\psi \coloneqq (\tens{x},\tens{\chi})}$ and define the microscopic action
\begin{equation}
	S_\psi[\psi] \coloneqq \tens{\chi \contr E}[\tens{x}] \,.
\end{equation}
Like the Hamiltonian, this action splits into a free and an interacting part,
$S_\psi = S_{\psi,0} + S_{\psi,\I}$, given by
\begin{align}
	S_{\psi,0}[\psi] &= \tens{\chi \contr} \, \bigl(\tens{\dot{x}} - \tens{\J \, \nabla_x} \H_0[\tens{x}]\bigr) \,,
	\label{micro:free_part_of_the_action} \\
	S_{\psi,\I}[\psi] &= \tens{\chi_p \contr \nabla_q} \H_\I[\tens{q}] \,.
	\label{eq:micro:interacting_part_of_the_action}
\end{align}
Then, the partition function \eqref{eq:micro:delta_trick_in_microscopic_path_integral} reads
\begin{equation}
	Z = \IC \is \bpi{\psi}{\tens{x^\ini}}{} \, \e^{\ii S_{\psi,0}[\psi] + \ii S_{\psi,\I}[\psi]} \,.
	\label{eq:micro:path_integral_in_compact_form}
\end{equation}

\subsection{Collective fields}
\label{sec:micro:collective_fields}
Usually, one is not interested in the microscopic state of the system but rather in its macroscopic properties. For this purpose, it is useful to introduce some collective fields $\Phi[\tens{\psi}]$ whose cumulants, i.\,e.~connected correlation functions, will contain the desired macroscopic information. Following \cite{fabis_kinetic_2018}, we choose $\Phi=\bigl(\Phi_f,\Phi_B\bigr)$ with $\Phi_f$ being the Klimontovich phase-space density and $\Phi_B$ the phase-space response field which encodes how the particle momenta are changed by a given interaction potential. Working in the Fourier space conjugate to phase-space, with ${\vect{s}=\bigl(\vect{k},\vect{l}\;\bigr)}$ denoting the Fourier vector conjugate to ${\vect{x}=\bigl(\vect{q},\vect{p}\bigr)}$, these two collective fields are given by
\begin{alignat}{2}
	\Phi_f\bigl(\vect{s}, t\bigr) &= \umi{6}{x} \, \e^{-\ii \vect{s} \cdot \vect{x}} \, \sum_{j=1}^N \, \dirac\bigl(\vect{x}-\vect{x}_j(t)\bigr) &&= \sum_{j=1}^N \, \e^{-\ii \vect{s} \cdot \vect{x}_j(t)} \,,
	\label{eq:micro:collective_f-field_in_Fourier_space} \\
	\Phi_B\bigl(\vect{s}, t\bigr) &= \umi{6}{x} \, \e^{-\ii \vect{s} \cdot \vect{x}} \, \sum_{j=1}^N \, \vect{\chi}_{p_j}(t) \cdot \vect{\nabla}_q \, \dirac\bigl(\vect{x}-\vect{x}_j(t)\bigr) &&= \sum_{j=1}^N \, \vect{\chi}_{p_j}(t) \cdot \ii \vect{k} \, \e^{-\ii \vect{s} \cdot \vect{x}_j(t)} \,.
	\label{eq:micro:collective_B-field_in_Fourier_space}
\end{alignat}

They allow us to express the interacting part of the action defined in \eqref{eq:micro:interacting_part_of_the_action} as
\begin{equation}
	S_{\psi,\I}[\tens{\psi}] = \usi{1} \usi{2} \, \Phi_f(-1) \, \sigfB(1,-2) \, \Phi_B(2) \,,
	\label{eq:micro:interacting_part_of_the_action_in_terms_of_collective_fields}
\end{equation}
where we used the conventional abbreviations ${(\pm r) \coloneqq \bigl(\pm\vect{s}_r, t_r\bigr)}$ as well as
\begin{equation}
	\usi{r} \coloneqq \fmi{6}{s_r} \bsiexp{t_r}{t_\ii}{\infty} \,,	
\end{equation}
and defined the interaction matrix element
\begin{equation}
	\sigfB(1,2) = \sigBf(2,1) \coloneqq - \, v\bigl(k_1, t_1\bigr) \, (2\pi)^9 \dirac\bigl(\vect{k}_1+\vect{k}_2\bigr) \, \dirac\bigl(\vect{l}_1\bigr) \, \dirac\bigl(\vect{l}_2\bigr) \, \dirac(t_1-t_2) \,.
	\label{eq:micro:interaction_matrix_element}
\end{equation}
The delta functions of $\vect{l}_1$ and $\vect{l}_2$ appearing in \eqref{eq:micro:interaction_matrix_element} are a consequence of the interaction potential being independent of the particle momenta.

As we will frequently encounter integrals over 1- and 2-point functions similar to those in \eqref{eq:micro:interacting_part_of_the_action_in_terms_of_collective_fields}, introducing a short-hand notation for these will greatly improve the clarity of the calculations. For general 1-point functions $A_1(1)$, $B_1(1)$ and 2-point functions $A_2(1,2)$, $B_2(1,2)$ we thus define their dot products as
\begin{alignat}{2}
	A_1 \cdot B_1 &\coloneqq \usi{\bar{1}} \, A_1(-\bar{1}) \, B_1(\bar{1}) \,,& \quad
	(A_2 \cdot B_2)(1,2) &\coloneqq \usi{\bar{1}} \, A_2(1,-\bar{1}) \, B_2(\bar{1},2) \,, \\
	(A_1 \cdot B_2)(1) &\coloneqq \usi{\bar{1}} \, A_1(-\bar{1}) \, B_2(\bar{1},1) \,,& \qquad
	(A_2 \cdot B_1)(1) &\coloneqq \usi{\bar{1}} \, A_2(1,-\bar{1}) \, B_1(\bar{1}) \,.
\end{alignat}
Using this notation, \eqref{eq:micro:interacting_part_of_the_action_in_terms_of_collective_fields} condenses to $\Phi_f \contr \sigfB \contr \Phi_B$.

Finally, we construct the generating functional of collective-field correlators by introducing a source field ${H = \bigl(H_f, H_B\bigr)}$ for the combined collective field $\Phi$ into the partition function \eqref{eq:micro:path_integral_in_compact_form},
\begin{equation}
	\Zcoll[H] \coloneqq \IC \bpi{\tens{\psi}}{\tens{x^\ini}}{} \, \e^{\ii S_0[\tens{\psi}] +\ii \Phi_f[\tens{\psi}] \contr \sigfB \contr \Phi_B[\tens{\psi}] + \ii H \contr \Phi[\tens{\psi}]} \,.
	\label{eq:micro:generating_functional_of_collective_field_cumulants}
\end{equation}
Functional derivatives of its logarithm $\Wcoll[H] \coloneqq \ln{\Zcoll[H]}$ with respect to the source field $H$, evaluated at $H=0$, yield the collective-field cumulants,
\begin{align}
	\G{f \dotsm f B \dotsm B}(1, \dotsc, n_f, 1', \dotsc, n'_B) &\coloneqq \bigl\langle\Phi_f(1) \dotsm \Phi_f(n_f) \, \Phi_B(1') \dotsm \Phi_B(n'_B)\bigr\rangle_\C \\
	&\,= \left. \prod_{u=1}^{n_f} \biggl(\der{H_f(u)}\biggr) \, \prod_{r=1}^{n_B} \biggl(\der{H_B(r')}\biggr) \, \Wcoll[H] \, \right|_{H=0} \,.
	\label{eq:micro:full_collective_n-point_cumulants}
\end{align}

Any macroscopic physical observable of the system can be derived from the pure phase-space density cumulants $\G{f \dotsm f}$. Of particular interest are their momentum moments. In real space, these can be computed by multiplying an $n_f$-point phase-space density cumulant with an appropriate function of the momenta, $F_p(\vect{p}_1, \dotsc, \vect{p}_{n_f})$, before integrating over these arguments. In Fourier space, this translates to applying the derivative operator
\begin{equation}
	\hat{F}_p \coloneqq \left. F_p\biggl(\ii \frac{\partial}{\partial \vect{l}_1}, \dotsc,\ii \frac{\partial}{\partial \vect{l}_{n_f}}\biggr) \, \right|_{\vect{l}_1=\dotsb=\vect{l}_{n_f}=0} \,.
	\label{eq:micro:momemtum_moment_operator}
\end{equation}
In the simple case $F_p=1$ this reduces a phase-space density cumulant to a cumulant of the spatial density field $\Phi_\rho$,
\begin{equation}
	\G{\rho \dotsm \rho}(\vect{k}_1, t_1, \dotsc,\vect{k}_{n_f}, t_{n_f}) = \G{f \dotsm f}(1,\dotsc,n_f) \, \Bigr|_{\vect{l}_1=\dotsb=\vect{l}_{n_f}=0} \,.
	\label{eq:micro:spatial_collective_field_cumulants}
\end{equation}

\subsection{Microscopic perturbation theory}
\label{sec:micro:free_cumulants}
For interacting particles, the collective-field cumulants can generally only be computed perturbatively. This assumes that the non-interacting theory, described by the free generating functional
\begin{equation}
	\fZcoll[H] \coloneqq \IC \bpi{\tens{\psi}}{\tens{x^\ini}}{} \, \e^{\ii S_0[\tens{\psi}] + \ii H \contr \Phi[\tens{\psi}]} \,,
\end{equation}
is exactly solvable in the sense that the free collective-field cumulants
\begin{equation}
	\fG_{f \dotsm fB \dotsm B}(1, \dotsc, n_f, 1', \dotsm, n'_B) = \left. \prod_{u=1}^{n_f} \biggl(\der{H_f(u)}\biggr) \, \prod_{r=1}^{n_B} \biggl(\der{H_B(r')}\biggr) \, \fWcoll[H] \, \right|_{H=0}
	\label{eq:micro:free_collective_n-point_cumulants}
\end{equation}
are exactly known. Here, we defined $\fWcoll[H] \coloneqq \ln{\fZcoll[H]}$.

If this is the case, one possible way to proceed is to pull the interacting part of the action in front of the path integral by replacing its $\Phi$-dependence with functional derivatives with respect to $H$, acting on the free generating functional $\fZcoll[H]$,
\begin{equation}
	\Zcoll[H] = \e^{\ii \der{H_f} \contr \sigfB \contr \der{H_B}} \, \fZcoll[H] \,.
\end{equation}
Expanding the exponential then yields a perturbative series that allows to calculate the interacting collective-field cumulants up to a finite order in $\sigfB$, and hence in the interaction potential $v$, only requiring the computation of a finite number of free cumulants. This approach has been discussed in detail in \cite{bartelmann_microscopic_2016,bartelmann_kinetic_2017} and we will refer to it as the \emph{microscopic perturbation theory}.

The reformulation of KFT that we will present in \secref{sec:macro} also requires the knowledge of the free collective-field cumulants. However, for that purpose it will prove more natural to work with free cumulants involving the \emph{dressed response field}
\begin{equation}
	\Phi_\F(1) \coloneqq (\sigfB \cdot \Phi_B)(1) = \usi{\bar{1}} \sigfB(1,-\bar{1}) \, \Phi_B(\bar{1})
\end{equation}
rather than the bare response field $\Phi_B$. The reason for this is that in the calculation of pure phase-space density cumulants $G_{f \dotsm f}$ every occurrence of the field $\Phi_B$ will be dressed with an interaction matrix element $\sigfB$ anyway. This is because $\Phi_B$ appears in the generating functional \eqref{eq:micro:generating_functional_of_collective_field_cumulants} only via the interaction term
\begin{equation}
	S_{\psi,\I}[\psi] = \Phi_f \contr \sigfB \contr \Phi_B = \Phi_f \contr \Phi_\F
	\label{eq:micro:interacting_part_of_the_action_in_terms_of_dressed_collective_fields_shorthand_notation}
\end{equation}
once we set $H_B$ to zero.

Let us thus define the dressed combined collective field $\tilde{\Phi} \coloneqq \bigl(\Phi_f,\Phi_\F\bigr)$ and its source field $\tilde{H} \coloneqq \bigl(H_f,H_\F\bigr)$ to construct the generating functional of dressed free collective-field correlators,
\begin{equation}
	\fdZcoll\bigl[\tilde{H}\bigr] \coloneqq \IC \bpi{\psi}{\tens{x^\ini}}{} \, \e^{\ii S_{\psi,0}[\psi] + \ii \tilde{H} \contr \tilde{\Phi}[\psi]} \,.
	\label{eq:micro:generating_functional_of_dressed_free_collective_field_correlators}
\end{equation}
The respective dressed free cumulants can then either be obtained from the cumulant-generating functional $\fdWcoll\bigl[\tilde{H}\bigr] \coloneqq \ln{\fdZcoll\bigl[\tilde{H}\bigr]}$ or via their relation to the bare cumulants,
\begin{align}
	\fG_{f \dotsm f \F \dotsm \F}(1, \dotsc, n_f, 1', &\dotsc, n'_\F) \coloneqq \prod_{u=1}^{n_f} \biggl(\der{H_f(u)}\biggr) \, \left. \prod_{r=1}^{n_\F} \biggl(\der{H_\F(r')}\biggr) \, \fdWcoll\bigl[\tilde{H}\bigr] \, \right|_{\tilde{H}=0}
	\label{eq:micro:free_dressed_collective_n-point_cumulants_from_generating_functional} \\
	&= \prod_{r=1}^{n_\F} \biggl(\usi{\bar{r}} \, \sigfB(r',-\bar{r})\biggr) \, \fG_{f \dotsm f B \dotsm B}(1,\dotsc,n_f,\bar{1},\dotsc,\bar{n}_\F) \,.
	\label{eq:micro:free_dressed_collective_n-point_cumulants_via_bare_cumulants}
\end{align}

In the case of initially correlated particles, the explicit computation of the free cumulants, be it dressed or bare, is rather involved. A thorough treatment of this can be found in \cite{fabis_kinetic_2018}, where a diagrammatic approach inspired by the Mayer cluster expansion \cite{mayer_molecular_1941} was developed to compute them systematically. For the considerations in the present work, though, it is completely sufficient to be aware of the free cumulants' physical interpretation and a few of their properties derived in \cite{fabis_kinetic_2018}, which we will only
summarise here:
\begin{enumerate}
	\item{
		\label{it:micro:macroscopic_meaning_of_free_cumulants}
		The cumulant $\fG_{f \dotsm f}(1,\dotsc,n_f)$ is the connected $n_f$-point correlation of the phase-space density observed at times $t_{1}, \dotsc, t_{n_f}$ that emerges from the ensemble average over the initially correlated, freely evolving particles.
		
		The cumulant $\fG_{f \dotsm f \F \dotsm \F}(1,\dotsc,n_f,$ $1',\dotsc,n'_\F)$, on the other hand, is the $n_\F$-th order	response 	of this $n_f$-point correlation to perturbations of the non-interacting system at times $t_{1'}, \dotsc, t_{n'_\F}$. If $n_\F = 1$, this is a linear response. Microscopically, these perturbations correspond to particles being deflected from their free trajectories by the forces $- \vect{\nabla} v$ acting between them and other particles. In Fourier space, this leads to the proportionality
		\begin{equation}
			\fG_{f \dotsm f \F \dotsm \F}(1,\dotsc,n_f,1',\dotsc,n'_\F) \propto - \ii \vect{k}_{r'} \, v(k_{r'}, t_{r'}) \, \dirac\bigl(\vect{l}_{r'}\bigr) \quad \forall \; r' \in \{1',\dotsc,n'_\F\} \,.
			\label{eq:micro:potential_gradients_and_delta_functions_of_l_in_free_cumulants}
		\end{equation}
		}
	\item{
		\label{it:micro:causality_of_free_cumulants}
		Causality tells us that the phase-space density $\Phi_f(u)$ evaluated at some time $t_u$ can only respond to perturbations experienced at earlier times $t_{r'} \leq t_u$. A more detailed analysis further shows that the spatial density ${\Phi_\rho\bigl(\vect{k}_u,t_u\bigr) = \Phi_f(u) \bigr|_{\vect{l}_u=0}}$ can in fact only respond to perturbations experienced at strictly earlier times $t_{r'} < t_u$. On the level of the free cumulants this manifests itself in the property
		\begin{equation}
			\begin{split}
				&\fG_{f \dotsm f \F \dotsm \F}(1,\dotsc,n_f,1',\dotsc,n'_\F) = 0 \quad \text{if} \quad \exists \; r' \in \{1',\dotsc,n'_\F\} \\
				&\text{such that} \;\; \bigl(t_{r'} > t_u\bigr) \;\; \text{or} \;\; \bigl(t_{r'} = t_u \;\text{and}\; \vect{l}_u = 0\bigr) \;\
				\; \forall \; u \in \{1,\dotsc,n_f\} \,.
			\end{split}
			\label{eq:micro:causal_structure_of_free_cumulant}
		\end{equation}
		In particular, every free pure $\Phi_\F$-cumulant vanishes identically,
		\begin{equation}
			\fG_{\F \dotsm \F}(1',\dotsc,n'_\F) = 0 \,.
			\label{eq:micro:free_pure_response_field_cumulant_vanishes}
		\end{equation}
		}
	\item{
		\label{it:micro:free_cumulants_in_statistically_homogeneous_systems}
		In statistically homogeneous systems, a free collective-field cumulant containing $n_f$ phase-space density fields is given by a sum of individual free $\ell$-particle cumulants with $1 \leq \ell \leq n_f$,
		\begin{equation}
			\fG_{f \dotsm f \F \dotsm \F}(1,\dotsc,n_f,1',\dotsc,n'_\F) = \sum_{\ell=1}^{n_f} G_{f \dotsm f \F \dotsm \F}^{(0,\ell)}(1,\dotsc,n_f,1',\dotsc,n'_\F) \,,
		\end{equation} 
		describing the contribution from correlations between $\ell$ particles. They satisfy
		\begin{equation}
			G_{f \dotsm f \F \dotsm \F}^{(0,\ell)}(1,\dotsc,n_f,1',\dotsc,n'_\F) \propto \bar{\rho}^\ell \, (2\pi)^3 \dirac\bigl(\vect{k}_1 + \dotsb + \vect{k}_{n_f} + \vect{k}_{1'} + \dotsb + \vect{k}_{n'_\F}\bigr) \,,
			\label{eq:micro:conservation_of_spatial_fourier_modes_in_free_cumulant}
		\end{equation}
		where $\bar{\rho}$ denotes the constant mean number density and the delta function arises due to the translational invariance in statistically homogeneous systems. Note that all $\ell$-particle cumulants with $\ell < n_f$ describe shot-noise contributions, arising due to the discrete nature of the density field. In the thermodynamic limit $N \rightarrow \infty$ these become negligible relative to the dominant term proportional to $\bar{\rho}^{n_f}$.
		}
\end{enumerate}

In \appref{app:free_cumulants_in_statistically_homogeneous_systems} we exemplarily list the resulting general expressions for the free 1- and 2-point cumulants in the case of statistically homogeneous and isotropic systems with Gaussian initial correlations.

\section{Resummed KFT}
\label{sec:macro}
From the perspective of most quantum and statistical field theories, the KFT partition function \eqref{eq:micro:path_integral_in_compact_form} is rather unusual in the sense that the path integral is expressed in terms of the microscopic fields $\psi$ even though we are actually interested in macroscopic fields like the phase-space density. To facilitate the application of already established perturbative as well as non-perturbative field-theoretical techniques, we would thus like to reformulate the KFT partition function as a path integral directly over the macroscopic fields we are interested in. This can be achieved by exploiting the exact solvability of the free theory and the fact that the interacting part of the action \eqref{eq:micro:interacting_part_of_the_action_in_terms_of_dressed_collective_fields_shorthand_notation} depends on $\psi$ only implicitly via the collective fields $\Phi_f[\psi]$ and $\Phi_\F[\psi]$.

\subsection{Macroscopic action}
\label{sec:macro:macroscopic_action}
We begin by introducing a functional delta distribution to replace the explicitly $\psi$-dependent field $\Phi_f[\psi]$ with a new formally $\psi$-independent field $f$,
\begin{equation}
	Z = \IC \bpi{\psi}{\tens{x^\ini}}{} \upi{f} \, \e^{\ii S_{\psi,0}[\psi] + \ii f \contr \Phi_\F[\psi]} \dirac\bigl[f - \Phi_f[\psi]\bigr] \,.
	\label{eq:macro:introducing_macroscopic_fields_into_the_path_integral}
\end{equation}
Note that this new field $f$ effectively still carries all the information of $\Phi_f[\psi]$ due to the delta distribution. Most importantly, $f$- and $\Phi_f$-correlation functions are precisely the same. But to emphasise the different origin we will deliberately call $f$ the macroscopic and $\Phi_f$ the collective phase-space density field in the following.

Similar to what we did in \eqref{eq:micro:delta_trick_in_microscopic_path_integral}, we can now express the delta distribution as a functional Fourier transform with respect to a new macroscopic auxiliary field $\beta$ conjugate to $f$,
\begin{equation}
	Z = \IC \bpi{\psi}{\tens{x^\ini}}{} \upi{f} \upi{\beta} \, \e^{\ii S_{\psi,0}[\psi] + \ii f \contr \Phi_\F[\psi] - \ii \beta \contr (f - \Phi_f[\psi])} \,.
	\label{eq:macro:introducing_macroscopic_auxiliary_field_into_the_path_integral}
\end{equation}
Pulling all $\psi$-independent parts to the front, we find that the remaining microscopic part of the path integral precisely takes the form of the free generating functional $\fdZcoll$,
\begin{align}
	Z	&= \upi{f} \upi{\beta} \e^{- \ii \beta \contr f} \IC \bpi{\psi}{\tens{x^\ini}}{} \, \e^{\ii S_{\psi,0}[\psi] + \ii \beta \contr \Phi_f[\psi] + \ii f \contr \Phi_\F[\psi]} \\
		&= \upi{\phi} \, \e^{- \ii \beta \contr f} \fdZcoll\bigl[\tilde{\phi}\bigr] \,,
	\label{eq:macro:separation_of_microscopic_and_macroscopic_part_of_the_path_integral}
\end{align}
where we defined the combined macroscopic fields $\phi \coloneqq (f,\beta)$ and $\tilde{\phi} = (\beta,f)$. It is $\tilde{\phi}$ which now plays the role of the dressed collective source field $\tilde{H}$ introduced in \secref{sec:micro:free_cumulants}.

Finally, using $\fdWcoll = \ln \fdZcoll$, we arrive at our desired result,
\begin{equation}
	Z = \upi{\phi} \, \e^{\ii S_\phi[\phi]}
	\label{eq:macro:macroscopic_path_integral_in_compact_form}
\end{equation}
with the macroscopic action
\begin{equation}
	S_\phi[\phi] \coloneqq - f \contr \beta - \ii \fdWcoll\bigl[\tilde{\phi}\bigr] \,.
	\label{eq:macro:macroscopic_action}
\end{equation}
We want to emphasise that this reformulation is \emph{exact} and hence \eqref{eq:macro:macroscopic_path_integral_in_compact_form} still contains the complete information on the microscopic dynamics, even though $S_\phi$ does not depend on $\psi$ any more. The microscopic information is now encoded in the free generating functional $\fdWcoll\bigl[\tilde{\phi}\bigr]$ and thus, by means of a functional Taylor expansion, in the dressed free collective-field cumulants,
\begin{align}
	\begin{split}
		\fdWcoll\bigl[\tilde{\phi}\bigr]
		= \!\!\! \sum_{n_\beta, n_f=0}^\infty \frac{1}{n_\beta! \, n_f!} \, &\prod_{u=1}^{n_\beta} \biggl(\usi{u} \beta(-u) \frac{\updelta}{\updelta H_f(u)}\biggr) \\
		\times &\prod_{r=1}^{n_f} \left. \biggl(\usi{r'} f(-r') \frac{\updelta}{\updelta H_\F(r')}\biggr) \, \fdWcoll\bigl[\tilde{H}\bigr] \, \right|_{\tilde{H}=0}
	\end{split} \\
	\begin{split}
		= \!\!\! \sum_{n_\beta, n_f=0}^\infty \frac{\ii^{n_\beta+n_f}}{n_\beta! \, n_f!} \, &\prod_{u=1}^{n_\beta} \biggl(\usi{u} \beta(-u)\biggr) \, \prod_{r=1}^{n_f} \biggl(\usi{r'} f(-r')\biggr) \\
		&\times \fG_{f \dotsm f \F \dotsm \F}(1,\dotsc,n_f,1',\dotsc,n'_\F) \,.
	\end{split}
	\label{eq:macro:expansion_of_W_0}
\end{align}

\subsection{Macroscopic perturbation theory}
\label{sec:macro:propagator_and_vertices}
The path integral in the form of \eqref{eq:macro:macroscopic_path_integral_in_compact_form} allows us to set up a new perturbative approach to KFT following the standard procedure in quantum and statistical field theory, i.\,e.~in terms of propagators and vertices. For this purpose, we first split up the action \eqref{eq:macro:macroscopic_action} into a propagator part $S_\Prop$, collecting all terms quadratic in $\phi$, and a vertex part $S_\Vert$, containing the remaining terms,
\begin{equation}
	S_\phi[\phi] \stackrel{!}{=} S_\Prop[\phi] + S_\Vert[\phi]
	\label{eq:macro:separating_macroscopic_action_into_propagator_and_vertex_parts}
\end{equation}
with
\begin{align}
	\ii S_\Prop[\phi] &\coloneqq - \frac{1}{2} \, \usi{1} \usi{2} \, \phi(-1) \, \iProp(1,2) \, \phi(-2) \,,
	\label{eq:macro:propagator_part_of_the_macroscopic_action} \\
	\begin{split}
		\ii S_\Vert[\phi] &\coloneqq \!\!\! \sum_{\substack{n_\beta, n_f=0 \\ n_\beta+n_f \neq 2}}^\infty \! \frac{1}{n_\beta! \, n_f!} \, \prod_{u=1}^{n_\beta} \biggl(\usi{u} \beta(-u)\biggr) \, \prod_{r=1}^{n_f} \biggl(\usi{r'} f(-r')\biggr) \\
		&\hphantom{\coloneqq \!\!\! \sum_{\substack{n_\beta, n_f=0 \\ n_\beta+n_f \neq 2}}^\infty \! \frac{1}{n_\beta! \, n_f!} \,} \times \Vert_{\beta \dotsm \beta f \dotsm f}(1,\dotsc,n_\beta,1',\dotsc,n'_f) \,,
	\end{split}
	\label{eq:macro:vertex_part_of_the_macroscopic_action}
\end{align}
introducing the inverse macroscopic propagator $\iProp$ and the macroscopic $(n_\beta+n_f)$-point vertices $\Vert_{\beta \dotsm \beta f \dotsm f}$.

Furthermore, we define the macroscopic generating functional $Z_\phi$ by introducing a source field ${M  = \bigl(M_f, M_\beta\bigr)}$ for the combined macroscopic field $\phi$ into the partition function,
\begin{equation}
	Z_\phi[M] \coloneqq \upi{\phi} \, \e^{\ii S_\phi[\phi] + \ii M \contr \phi} \,.
\end{equation}
Then, the vertex part of the action can be pulled in front of the path integral by replacing its $\phi$-dependence with functional derivatives with respect to $M$, $\hat{S}_\Vert \coloneqq S_\Vert\bigl[\der{M}\bigr]$, acting on the remaining path integral,
\begin{align}
	Z_\phi[M]	&= \e^{\ii \hat{S}_\Vert} \upi{\phi} \, \e^{- \frac{1}{2} \phi \contr \iProp \contr \phi + \ii M \contr \phi} \\
							&= \e^{\ii \hat{S}_\Vert} \, \e^{\frac{1}{2} (\ii M) \contr \Prop \contr (\ii M)} \,.
							\label{eq:macro:path_integral_split_into_vertex_and_propagator_part}
\end{align}
In the last step we dropped a constant functional determinant, as it can be absorbed into the normalization of the path integral. Expanding the first exponential in \eqref{eq:macro:path_integral_split_into_vertex_and_propagator_part} in orders of the vertices now gives rise to a new perturbative approach that we will refer to as the \emph{macroscopic perturbation theory}.

Within this approach, the interacting macroscopic-field cumulants are obtained via
\begin{equation}
	G_{f \dotsm f \beta \dotsm \beta}(1, \dotsc, n_f, 1', \dotsm, n'_\beta) = \left. \prod_{u=1}^{n_f} \biggl(\der{M_f(u)}\biggr) \, \prod_{r=1}^{n_\beta} \biggl(\der{M_\beta(r')}\biggr) \, \Wmac[M] \, \right|_{M=0} \,,
	\label{eq:KFT:macro:full_macroscopic_n-point_cumulants}
\end{equation}
where we defined the macroscopic cumulant-generating functional $\Wmac[M] \coloneqq \ln{\Zmac[M]}$. In particular, this implies that the lowest perturbative order of the 2-point phase-space density cumulant $\Gff$ is given by the $ff$-component of the macroscopic propagator,
\begin{align}
	\Gff(1,2) &= \left. \der{M_f(1)} \, \der{M_f(2)} \, \Wmac[M] \, \right|_{M=0} \\
	&= \Propff(1,2) + \text{terms involving vertices} \,.
	\label{eq:macro:ff-propagator_is_tree-level_result_of_ff-cumulant}
\end{align}
For a systematic computation of the higher-order contributions involving vertices, we first need to specify an explicit expansion scheme. We will return to this at the end of \secref{sec:macro:diagrams} after investigating the general properties of the macroscopic perturbation theory in more detail first.

To find explicit expressions for $\iProp$ and $\Vert_{\beta \dotsm \beta f \dotsm f}$, we insert \eqref{eq:macro:expansion_of_W_0} into \eqref{eq:macro:macroscopic_action} and identify terms with \eqref{eq:macro:propagator_part_of_the_macroscopic_action} and \eqref{eq:macro:vertex_part_of_the_macroscopic_action}, respectively,
\begin{align}
	\iProp(1,2)	=
	\begin{pmatrix}
		\iPropff & \; \iPropfb \\
		\iPropbf & \; \iPropbb
	\end{pmatrix} \! (1,2)
	&=
	\begin{pmatrix}
		\fGFF & \; \ii \, \id + \fGFf \\
		\ii \, \id + \fGfF & \; \fGff
	\end{pmatrix} \! (1,2) \,,
	\label{eq:macro:inverse_propagator} \\
	\Vert_{\beta \dotsm \beta f \dotsm f}(1,\dotsc,n_\beta,1',\dotsc,n_f') &= \ii^{n_\beta+n_f} \, \fG_{f \dotsm f \F \dotsm \F}(1,\dotsc,n_\beta,1',\dotsc,n'_f) \,.
	\label{eq:macro:vertices}
\end{align}
Here, $\id$ denotes the identity 2-point function,
\begin{equation}
	\id(1,2) \coloneqq (2\pi)^3 \dirac\bigl(\vect{k}_1+\vect{k_2}\bigr) \, (2\pi)^3 \dirac\bigl(\vect{l}_1+\vect{l_2}\bigr) \, \dirac(t_1-t_2) \,.
	\label{eq:macro:identity_2-point_function}
\end{equation}
Note that every $\beta$-component of an inverse propagator or vertex in \eqref{eq:macro:inverse_propagator} and \eqref{eq:macro:vertices} corresponds to an $f$-component of a free cumulant, whereas every $f$-component of an inverse propagator or vertex corresponds to an $\F$-component of a free cumulant. Together with \eqref{eq:micro:free_pure_response_field_cumulant_vanishes}, this directly allows us to conclude that $\iPropff$ and $\Vert_{f \dotsm f}$ vanish identically.

The propagator $\Prop$ is then obtained by a combined matrix and functional inversion of \eqref{eq:macro:inverse_propagator}, defined via the following matrix integral equation,
\begin{equation}
	\usi{\bar{1}} \, \Prop(1,-\bar{1}) \; \iProp(\bar{1},2) \stackrel{!}{=} \id(1,2) \; \id_2 \,.
\end{equation}
The matrix part of this inversion can be performed immediately and yields
\begin{equation}
	\Prop(1,2) =
	\begin{pmatrix}
		\Propff & \Propfb \\
		\Propbf & \Propbb
	\end{pmatrix} \! (1,2)
	=
	\begin{pmatrix}
		\PropR \contr \fGff \contr \PropA & \quad - \ii \PropR \\
		- \ii \PropA & \quad 0
	\end{pmatrix} \! (1,2) \,,
	\label{eq:macro:explicit_expression_for_the_propagator}
\end{equation}
where we defined the abbreviations
\begin{equation}
	\PropR(1,2) = \PropA(2,1) \coloneqq \Bigl(\id - \ii \fGfF\Bigr)^{-1} (1,2)
	\label{eq:macro:retarded_and_advanced_propagator}
\end{equation}
for the remaining functional inverses. In \appref{app:computing_the_macroscopic_propagator} we describe how these as well as $\Propff$ can be computed explicitly for a given physical system.

But even without specifying the system it is always possible to express \eqref{eq:macro:retarded_and_advanced_propagator} formally in terms of a Neumann series by expanding the functional inverse in orders of $\ii \fGfF$,
\begin{align}
	\PropR(1,2) = \PropA(2,1)
		&\,= \sum_{n=0}^\infty \, \Bigl(\ii \fGfF\Bigr)^n(1,2)
		\label{eq:macro:expansion_of_retarded_and_advanced_propagator_with_powers} \\
		&\,= \id(1,2) + \ii \fGfF(1,2) + \usi{\bar{1}} \, \ii \fGfF(1,-\bar{1}) \, \ii \fGfF(\bar{1},2) + \dotsb \,.
		\label{eq:macro:expansion_of_retarded_and_advanced_propagator_with_integrals}
\end{align}
Recalling $\fGfF = \fGfB \contr \sigBf$ from \eqref{eq:micro:free_dressed_collective_n-point_cumulants_via_bare_cumulants}, we can see that $\PropR$ and $\PropA$ contain terms of arbitrarily high order in $\sigBf$ and hence in the interaction potential $v$. Inserting them into the $ff$-component of \eqref{eq:macro:explicit_expression_for_the_propagator} yields
\begin{equation}
	\Propff(1,2) = \sum_{n_\R,n_\A=0}^\infty \, \biggl(\Bigl(\ii \fGfB \contr \sigBf\Bigr)^{n_\R} \!\!\! \contr \fGff \contr \Bigl(\sigfB \contr \ii \fGBf\Bigr)^{n_\A}\biggr)(1,2) \,.
	\label{eq:macro:expansion_of_statistical_propagator_with_powers}
\end{equation}
This demonstrates that the lowest perturbative order within the macroscopic approach already captures some features which could only be described at infinitely high order within the microscopic perturbation theory. Consequently, we expect the macroscopic perturbation theory to be generally more powerful than the microscopic one.

We can make this statement more precise by recalling from \eqref{eq:macro:ff-propagator_is_tree-level_result_of_ff-cumulant} that $\Propff$ is exactly the result we get for $\Gff$ by taking no vertices into account. Due to \eqref{eq:macro:vertex_part_of_the_macroscopic_action} and \eqref{eq:macro:vertices} this is equivalent to computing $\Gff$ while ignoring all contributions involving free $n$-point cumulants with $n \neq 2$. Thus, the transition from the micro- to the macroscopic formulation leads to a resummation of all contributions appearing in the microscopic perturbative series that only contain free 2-point cumulants. In a statistically homogeneous system this corresponds to a resummation of all contributions which do not lead to any mode-coupling beyond the one already introduced by the free evolution. Due to this property, we will refer to the macroscopic reformulation of the KFT framework as \emph{resummed KFT} (RKFT).

To understand what this resummation corresponds to at the level of the microscopic particle dynamics, we need to take a closer look at the meaning of the free cumulants appearing in the propagator. From the properties \ref{it:micro:macroscopic_meaning_of_free_cumulants} and \ref{it:micro:causality_of_free_cumulants} discussed in \secref{sec:micro:free_cumulants} we know that $\ii \fGfF(1,2)$ describes the linear response of the phase-space density $\Phi_f(1)$ at time $t_1$ to particles being deflected from their free trajectories by some perturbing forces acting on them at an earlier time $t_2 \leq t_1$. The product $\ii \fGfF(1,-\bar{1}) \, \ii \fGfF(\bar{1},2)$, appearing under the integral in \eqref{eq:macro:expansion_of_retarded_and_advanced_propagator_with_integrals}, accordingly describes the linear response of $\Phi_f(1)$ to a situation in which the forces generated by those already deflected particles perturb the trajectories of so far still freely evolving particles at an intermediate time $t_{\bar{1}}$ with $t_2 \leq t_{\bar{1}} \leq t_1$.

\begin{figure}
	\centering
	\includegraphics{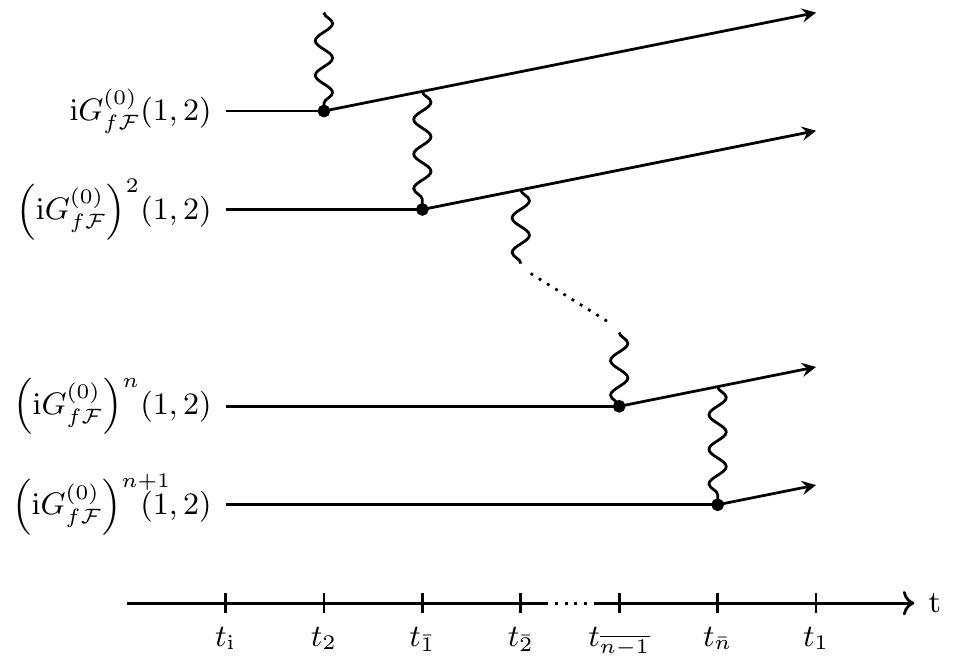}
	\caption{Illustration of the microscopic particle dynamical processes resummed in the retarded propagator $\PropR(1,2)$. The arrows represent particle trajectories while the wavy lines depict forces acting on those particles. The different contributions to $\PropR(1,2)$ denoted on the left side correspond to different truncations of an iterative process where in each step particles are deflected from their free trajectories by the forces generated by the already deflected particles from the previous step. The open wavy line at the top represents the forces initiating the whole process at time $t_2$.}
	\label{fig:macro:resummed_particle_dynamical_processes}
\end{figure}

This process can be repeated for any possible number of iterations $n$, with the deflections occurring continuously at all possible intermediate times $t_{\bar{1}}, \dotsc, t_{\bar{n}}$ with $t_2 \leq t_{\bar{1}} \leq \dotsb \leq t_{\bar{n}} \leq t_1$, as illustrated in \figref{fig:macro:resummed_particle_dynamical_processes}. $\PropR$ and $\PropA$ capture this by summing over all possible numbers of linear response cumulants $\ii \fGfF$ and integrating over all intermediate times. The correct time ordering under the integral is ensured by the property $\fGfF(1,2) \propto \step(t_1-t_2) \,,$ following from \eqref{eq:micro:causal_structure_of_free_cumulant}. Consequently, we find
\begin{equation}
	\PropR(1,2) = \PropA(2,1) = 0 \quad \text{if} \quad t_2 > t_1 \,,
	\label{eq:macro:causal_structure_of_retarded_and_advanced_propagator}
\end{equation}
and we will hence call $\PropR$ and $\PropA$ the \emph{retarded and advanced (macroscopic) propagators}, respectively. If we do not want to explicitly distinguish between the two, we will refer to them more generally as the \emph{causal (macroscopic) propagators}.

In $\Propff = \PropR \contr \fGff \contr \PropA$ the forces generated by the freely evolving correlated particles associated with $\fGff$ cause the perturbation needed to initiate the iterative deflection process encoded in $\PropR$ and $\PropA$. Accordingly, $\Propff$ describes the connected 2-point phase-space density correlation emerging from the ensemble average over initially correlated particles that have undergone this process. We will thus call $\Propff$ the \emph{statistical (macroscopic) propagator} --- adopting the nomenclature used in nonequilibrium quantum and statistical field theory.

We want to stress that even though $\PropR$ and $\PropA$ only contain linear response cumulants, $\Propff$ will generally contain contributions that are nonlinear in the initial phase-space density correlation, as the free-streaming of particles in $\fGff$ itself builds up those nonlinearities as long as the initial particle momenta are correlated \cite{fabis_kinetic_2018}.

\subsection{Feynman diagrams}
\label{sec:macro:diagrams}
Before applying the macroscopic perturbation theory to cosmic structure formation in \secref{sec:cosmo}, let us first examine some of its general properties. For this purpose, it is convenient to introduce diagrammatic representations for the propagators and vertices. We want to do this in such a way that the diagrams indicate the causal structure inherited from the free collective-field cumulants. Combining \eqref{eq:micro:causal_structure_of_free_cumulant} and \eqref{eq:macro:vertices}
yields
\begin{equation}
	\begin{split}
		&\Vert_{\beta \dotsm \beta f \dotsm f}(1,\dotsc,n_\beta,1',\dotsc,n'_f) = 0 \quad \text{if} \quad \exists \; r' \in \{1',\dotsc,n'_f\} \\
		&\text{such that} \;\; \bigl(t_{r'} > t_u\bigr) \;\; \text{or} \;\; \bigl(t_{r'} = t_u \;\text{and}\; \vect{l}_u = 0\bigr) \;\
		\; \forall \; u \in \{1,\dotsc,n_\beta\} \,,
	\end{split}
	\label{eq:macro:causal_structure_of_vertex}
\end{equation}
meaning that every $f$-argument of a vertex must always be evaluated at an earlier time than at least one of its $\beta$-arguments. We visualise this by placing incoming arrowheads on the $f$-legs and outgoing arrowheads on the $\beta$-legs, indicating the direction of time. A general vertex is then represented as
\begin{equation}
	\Vert_{\beta \dotsm \beta f \dotsm f}(1,\dotsc,n_\beta,1',\dotsc,n'_f) \cong \raisebox{-0.5\height+1ex}{\includegraphics{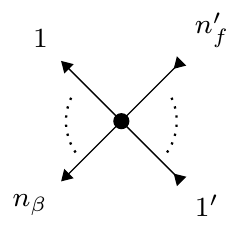}} \quad \text{if $n_\beta \geq 1$,}
	\label{eq:macro:diagrammatic_representation_of_vertices}
\end{equation}
where the dots indicate a leg to be drawn for each argument of the vertex. If $n_\beta=0$, it follows from \eqref{eq:micro:free_pure_response_field_cumulant_vanishes} and \eqref{eq:macro:vertices} that the vertex vanishes identically, $\Vert_{f \dotsm f} = 0$.

The propagator \eqref{eq:macro:explicit_expression_for_the_propagator} contains the building blocks $\fGff$, $\PropR$ and $\PropA$. For the first one of these we can directly adopt the representation \eqref{eq:macro:diagrammatic_representation_of_vertices} by means of $\fGff = - \Vert_{\beta \beta}$,
\begin{equation}
	- \fGff(1,2) \cong \raisebox{-0.6ex}{\includegraphics{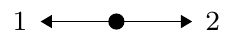}} \,.
	\label{eq:macro:diagrammatic_representation_of_free_fF_cumulant}
\end{equation}
The causal propagators $\PropR$ and $\PropA$, however, require a new kind of diagram. In accordance with their respective retarded or advanced causal structure we represent them as a line with an arrowhead in the middle, pointing towards the later time,
\begin{equation}
	-\ii \PropR(1,2) = -\ii \PropA(2,1) \cong \raisebox{-0.6ex}{\includegraphics{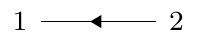}} \,,
	\label{eq:macro:diagrammatic_representation_of_retarded_and_advanced_propagator}
\end{equation}
additionally introducing a conventional factor $-\ii$. Whenever we connect the diagrams \eqref{eq:macro:diagrammatic_representation_of_vertices}, \eqref{eq:macro:diagrammatic_representation_of_free_fF_cumulant} and \eqref{eq:macro:diagrammatic_representation_of_retarded_and_advanced_propagator} in such a way that there are consecutive arrowheads on a line pointing into the same direction, we further join these arrowheads into one. This yields the following diagrammatic representations for the different components of the propagator,
\begin{equation}
	\Prop(1,2) \cong
	\begin{pmatrix}
		\raisebox{-0.6ex}{\includegraphics{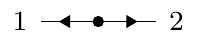}} & \raisebox{-0.6ex}{\includegraphics{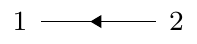}} \\
		\raisebox{-0.6ex}{\includegraphics{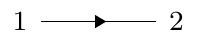}} & 0
	\end{pmatrix} \,.
	\label{eq:macro:diagrammatic_representation_of_propagator}
\end{equation}

Every term in the macroscopic perturbative expansion \eqref{eq:macro:path_integral_split_into_vertex_and_propagator_part} can now be represented by combining the diagrammatic expressions \eqref{eq:macro:diagrammatic_representation_of_vertices} and \eqref{eq:macro:diagrammatic_representation_of_propagator} for the vertices and propagators appearing in that term. Furthermore, it is a well-established fact in quantum and statistical field theory that the terms appearing in the cumulant-generating functional ${\Wmac[M] \coloneqq \ln{\Zmac[M]}}$ always correspond to connected diagrams. We will thus only consider those in the following.

In such a diagram every $f$-leg of each vertex will be attached to an $f$-end of a propagator and every $\beta$-leg of each vertex will be attached to a $\beta$-end of a propagator. This allows some immediate conclusions. First, one notices that the arrowhead directions of propagator-ends and vertex-legs being connected in this way will always agree. There will thus be a consistent and continuous \emph{time-flow} throughout the complete diagram. Second, the fact that there are no propagators or vertices with only incoming arrows implies that the time-flow has no sinks. In \appref{app:proofs_of_the_Feynman_rules:causality_rule} we demonstrate how these two properties can be used to derive the following
\begin{description}
	\item[Causality rule]{\label{it:macro:causality_rule}
	A diagram is causally forbidden and vanishes identically if it has only incoming arrows on its outer legs or contains a subdiagram which does so.}
\end{description}

If the system of interest is statistically homogeneous, the macroscopic perturbation theory has some additional very convenient properties. First of all, we recall from \eqref{eq:micro:conservation_of_spatial_fourier_modes_in_free_cumulant} that the free cumulants conserve spatial Fourier modes in this case. This property is translated to the propagator and vertices,
\begin{align}
	\Prop(1,2) & \propto (2\pi)^3 \dirac\bigl(\vect{k}_1+\vect{k}_2\bigr) \,,
	\label{eq:macro:conservation_of_spatial_fourier_modes_in_propagator} \\
	\Vert_{\beta \dotsm \beta f \dotsm f}(1, \dotsc, n_\beta, 1', \dotsc, n'_f) & \propto (2\pi)^3 \dirac\bigl(\vect{k}_1 + \dotsb + \vect{k}_{n_\beta} + \vect{k}_{1'} + \dotsb + \vect{k}_{n'_f} \bigr) \,,
	\label{eq:macro:conservation_of_spatial_fourier_modes_in_vertices}
\end{align}
as can be checked by inserting \eqref{eq:micro:interaction_matrix_element} and \eqref{eq:macro:identity_2-point_function} into \eqref{eq:macro:inverse_propagator} and \eqref{eq:macro:vertices}. Consequently, any connected diagram has to conserve spatial Fourier modes as well.

Furthermore, we show in \appref{app:proofs_of_the_Feynman_rules:homogeneity_rule} how this property allows to prove the following
\begin{description}
	\item[Homogeneity rule]{\label{it:macro:homogeneity_rule}
	In statistically homogeneous systems, a diagram vanishes identically if it contains a so-called tadpole subdiagram, i.\,e.~a subdiagram which is connected to the rest of the diagram solely via a single propagator.}
\end{description}
The physical interpretation of this rule is straightforward. The only information carried by a diagram with only one external leg is the value of the mean phase-space density. But a homogeneous background density can not affect the dynamics of the particles, as these depend on potential gradients only.

The two Feynman rules will enable us to drastically reduce the number of contributions to the macroscopic perturbative expansion that have to be taken into account. The Homogeneity rule further implies the applicability of the conventional loop expansion scheme when considering statistically homogeneous systems, as this rule ensures that there will always only be a finite number of non-vanishing diagrams with a given number of loops. The complete procedure for calculating the $L$-loop expression for an interacting $n_f$-point phase-space density cumulant of a statistically homogeneous system within the macroscopic perturbation theory is then as follows:
\begin{enumerate}
	\item{
		Draw $n_f$ points and label these with the Fourier space arguments 1 to $n_f$.
		}
	\item{
		By combining the propagators \eqref{eq:macro:diagrammatic_representation_of_propagator} and vertices \eqref{eq:macro:diagrammatic_representation_of_vertices} such that the arrow directions of joint legs always agree, draw all possible connected diagrams with $n_f$ outgoing external propagators ending at the labelled points that are allowed by the
		Causality as well as the Homogeneity rule and contain exactly $L$ loops.
		}
	\item{
		Divide each diagram by its respective symmetry factor, i.\,e.~the number of possible permutations of internal propagators and vertices which leave the labelled diagram invariant.
		}
	\item{
		Translate each diagram into its respective functional expression, taking into account that every link between a propagator and a vertex corresponds to a Fourier space integral over the argument at their joint legs, and sum up all resulting expressions.
		}
\end{enumerate}

In this sense, $\Propff$ is the 0-loop or --- adopting the usual field-theoretical nomenclature --- tree-level result of the 2-point phase-space density cumulant $\Gff$. Two examples of diagrams appearing in the 1-loop result of $\Gff$ are
\begin{align}
	\begin{split}
		\raisebox{-0.5\height+0.7ex}{\includegraphics{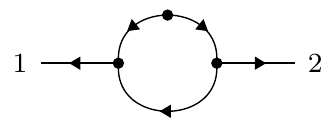}} \cong \half \int \! \d\bar{1} \dotsm \d\bar{6} \; &\Propfb(1, -\bar{1}) \, \Vert_{\beta f f}(\bar{1}, -\bar{2}, -\bar{3}) \, \Propff(\bar{2}, -\bar{4}) \\
		\times \, &\Propfb(\bar{3}, -\bar{5}) \, \Vert_{f \beta \beta}(\bar{4}, \bar{5}, -\bar{6}) \, \Propbf(\bar{6}, 2) \,,
	\end{split} \\
	\begin{split}
		\frac{1}{2} \raisebox{-0.5ex}{\includegraphics{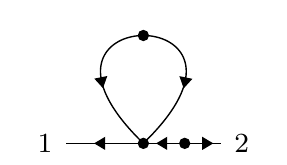}} \cong \frac{1}{2} \int \! \d\bar{1} \dotsm \d\bar{4} \; &\Propfb(1, -\bar{1}) \, \Vert_{\beta f f f}(\bar{1}, -\bar{2}, \bar{3}, -\bar{4}) \\
		\times \, &\Propff(\bar{2}, -\bar{3}) \, \Propff(\bar{4}, 2) \,.
	\end{split}
\end{align}

Having proven that for a statistically homogeneous system it is possible to systematically order the macroscopic perturbative contributions by their number of loops, it remains to be shown that this is also a physically motivated expansion scheme. Unfortunately, there is no obvious expansion parameter which would allow us to estimate the magnitude of the $L$-loop order contribution simply by the $L$-th power of said parameter.	However, we can nevertheless gain some intuition for the relevance of different perturbative terms by taking a closer look at the structure of
the macroscopic vertices.

From \eqref{eq:macro:vertices} and
\autoref{it:micro:macroscopic_meaning_of_free_cumulants} of \secref{sec:micro:free_cumulants} we know that a vertex $\Vert_{\beta \dotsm \beta f \dotsm f}$ with $n_\beta$ $\beta$-legs and $n_f$ $f$-legs describes the free $n_\beta$-point phase-space density cumulant and its $n_f$-th order response to interactions between the particles. Such a vertex introduces two distinct types of nonlinearities into the perturbative calculation: On the one hand, it acts as an $n_\beta$-point correlated source of the interaction potential, which is nonlinear in the initial phase-space density correlations if the initial particle momenta are correlated and $n_\beta \geq 2$ \cite{fabis_kinetic_2018}. On the other hand, the vertex describes a nonlinear response of the phase-space density to the particle interactions if $n_f \geq 2$. The Homogeneity rule further implies that only vertices with at least three legs contribute, $n_\beta + n_f \geq 3$, meaning that every single vertex will introduce at least one of those two types of nonlinearities.

Crucially, fixing the number of loops limits the maximal number of vertices appearing in a perturbative contribution as well as the maximal number of $\beta$- and $f$-legs that any of those vertices can have. The loop order thus characterises the overall degree of nonlinearity caused by interactions that are taken into account. Therefore, an expansion scheme in loop orders indeed seems to be a plausible perturbative approach for probing the nonlinear dynamics of an interacting system of classical particles. Further investigations of the validity and convergence of this scheme are still required, though, and will be subject of future work.

\section{Tree-level calculation of cosmic structure formation}
\label{sec:cosmo}
Having discussed the general formalism of RKFT, we will now use it specifically to describe the growth of cosmic structures in a standard $\Lambda$CDM cosmology. In this work, we consider the whole cosmic matter content to be made up of a single species of identical dark matter particles, and compute the macroscopic propagator to obtain the tree-level result of $\Gff$. Loop corrections as well as the evolution of a joint system of dark and baryonic matter will be investigated in separate papers.

\subsection{Newtonian dynamics of dark matter particles}
\label{sec:cosmo:dynamics_of_dark_matter_particles}
First, we need to specify our choice of coordinates. Instead of the cosmic time $t$ we use
\begin{equation}
	\eta(t) \coloneqq \log{\frac{D_+(t)}{D_+(t_\ii)}}
	\label{eq:cosmo:time_coordinate}
\end{equation}
as our time coordinate, following \cite{pietroni_coarse-grained_2012}, with $D_+$ being the usual linear growth factor. Note that this definition implies $\eta_\ii \coloneqq \eta(t_\ii) = 0$. The spatial coordinate is chosen to be comoving with the homogenous expansion of the background space-time, $\vect{q} \coloneqq \vect{r} / a$, where $\vect{r}$ denotes the physical coordinate and $a$ the cosmological scale factor. As momentum variable we use $\vect{p} \coloneqq \d \vect{q} / \d \eta$, i.\,e.~the comoving velocity with respect to the coordinate time $\eta$.

In \appref{app:equations_of_motion_for_a_dm_particle} we show that the resulting Newtonian equations of motion for the phase-space trajectory of a dark matter particle read
\begin{align}
	\td{\vect{q}}{\eta} &= \vect{p} \,,
	\label{eq:cosmo:EOM_of_DM_particle_position} \\
	\td{\vect{p}}{\eta} &= \biggl(1 - \frac{3}{2} \, \frac{\Omega_\m}{f_+^2}\biggr) \, \vect{p} - \vect{\nabla}_q \tilde{V} \approx - \frac{1}{2} \, \vect{p} - \vect{\nabla}_q \tilde{V} \,,
	\label{eq:cosmo:EOM_of_DM_particle_momentum}
\end{align}
with the dimensionless matter density parameter $\Omega_m$ defined in \eqref{eq:app:dimensionless_matter_density_parameter}, the growth function $f_+ \coloneqq \d \ln{D_+} / \d \ln{a}$ and the rescaled Newtonian gravitational potential $\tilde{V}$ satisfying the Poisson equation
\begin{equation}
	\vect{\nabla}_q^2 \tilde{V} = \frac{3}{2} \, \frac{\Omega_\m}{f_+^2} \, \frac{\Phi_\rho - \bar{\rho}}{\bar{\rho}} \approx \frac{3}{2} \, \frac{\Phi_\rho - \bar{\rho}}{\bar{\rho}} \,.
	\label{eq:cosmo:poisson_equation}
\end{equation}
Due to our choice of coordinates, $\bar{\rho} = \langle\Phi_\rho\rangle$ is the mean comoving number density of dark matter particles. We additionally used $\Omega_\m / f_+^2 \approx 1$, which is a very good approximation during the matter- and $\Lambda$-dominated cosmological epochs that we are interested in \cite{bernardeau_large-scale_2002,matarrese_resumming_2007}.

The free equations of motion, obtained by setting $\tilde{V} = 0$, can be written as
\begin{equation}
	\biggl(\td{}{\eta} + \mathcal{E}_0\biggr) \, \vect{x} = 0 \,, \qquad
	\mathcal{E}_0(\eta) =
	\begin{pmatrix}
		0_3 \;\; & - \id_3 \\
		0_3 \;\; & \frac{1}{2} \, \id_3
	\end{pmatrix} \,.
	\label{eq:app:free_dm_eom}
\end{equation}
Following \cite{bartelmann_trajectories_2015}, the corresponding single-particle retarded Green's function $\mathcal{G}$ defined in \eqref{eq:micro:retarded_Green's_function} is then given by
\begin{equation}
	\mathcal{G}(\eta,\eta') = \exp\left\{-\int_{\eta'}^\eta \d\bar{\eta} \, \mathcal{E}_0\right\} \, \step(\eta-\eta')
\end{equation}
and thus has the components
\begin{align}
	g_{qq}(\eta, \eta') &= \step(\eta - \eta') \,,
	\label{eq:cosmo:single_particle_qq_green's_function} \\
	g_{qp}(\eta, \eta') &= \step(\eta - \eta') \, 2 \, \bigl(1 - \e^{-\frac{1}{2} (\eta - \eta')}\bigr) \,,
	\label{eq:cosmo:single_particle_qp_green's_function} \\
	g_{pq}(\eta, \eta') &= 0 \,,
	\label{eq:cosmo:single_particle_pq_green's_function} \\
	g_{pp}(\eta, \eta') &= \step(\eta - \eta') \, \e^{-\frac{1}{2} (\eta - \eta')} \,.
	\label{eq:cosmo:single_particle_pp_green's_function}
\end{align}
Furthermore, the Poisson equation \eqref{eq:cosmo:poisson_equation} is solved by 
\begin{equation}
	\tilde{V}(\vect{q},t) = \sum_{j=1}^N \, v\bigl(|\vect{q}-\vect{q}_j(t)|\bigr) \,,
\end{equation}
with the single-particle gravitational potential reading
\begin{equation}
	v(k) = - \frac{3}{2} \, \frac{1}{\bar{\rho} \, k^2}
	\label{eq:cosmo:single_particle_potential}
\end{equation}
in Fourier space. We will use this potential in the definition \eqref{eq:micro:interaction_matrix_element} of the interaction matrix element $\sigfB$.

Now all that is left to do is to specify the initial conditions. We choose to fix the initial time to some instant early in the matter-dominated epoch when the cosmic density and momentum fields were still well-described by Gaussian random fields. For this case, an exact expression for the initial phase-space probability distribution $\mathcal{P}_\ii$ was derived in \cite{bartelmann_microscopic_2016}, depending only on the initial density contrast power spectrum $P_\delta^\ini(k) = P_\delta(k,\eta_\ii)$, where $\delta \coloneqq (\Phi_\rho - \bar{\rho}) / \bar{\rho}$
and
\begin{equation}
	(2\pi)^3 \dirac\bigl(\vect{k}_1 + \vect{k}_2\bigr) \, P_\delta(k_1, \eta) \coloneqq \bigl\langle\delta(\vect{k}_1, \eta) \, \delta(\vect{k}_2, \eta)\bigr\rangle = \frac{1}{\bar{\rho}^2} \, \Grr\bigl(\vect{k}_1, \eta, \vect{k}_2, \eta\bigr) \,.
	\label{eq:cosmo:density_contrast_power_spectrum}
\end{equation}

General expressions for the free collective-field cumulants resulting from this choice of initial conditions have been derived in \cite{fabis_kinetic_2018}, and in \appref{app:free_cumulants_in_statistically_homogeneous_systems} we list these for the 1- and 2-point cumulants used in this work. Inserting \eqref{eq:cosmo:single_particle_qq_green's_function}--\eqref{eq:cosmo:single_particle_pp_green's_function} and \eqref{eq:cosmo:single_particle_potential} into these expressions yields
\begin{align}
	\fGf(1) &= (2\pi)^3 \dirac\bigl(\vect{k}_1\bigr) \, \bar{\rho} \, \e^{-\frac{\sigma_p^2}{2} \cosmoT_1^2 l_1^2} \,,
	\label{eq:cosmo:free_f_cumulant} \\
	\begin{split}
		\fGfF(1,2) &= (2\pi)^3 \dirac\bigl(\vect{k}_1 + \vect{k}_2\bigr) \, (2\pi)^3 \dirac\bigl(\vect{l}_2\bigr) \, \step(\eta_1 - \eta_2) \\
		&\= \times\frac{- 3 \, \ii}{2} \, \biggl[2 - \biggl(2 - \frac{\vect{k}_1 \cdot \vect{l}_1}{k_1^2}\biggl) \, \cosmoT_{12}\biggr] \, \e^{- \frac{\sigma_p^2}{2} \bigl[2 \, (\cosmoT_2 - \cosmoT_1) \vect{k}_1 + \cosmoT_1 \vect{l}_1\bigr]^2} \,,
	\end{split}
	\label{eq:cosmo:free_fF_cumulant} \\
	\fGff(1,2) &= (2\pi)^3 \dirac\bigl(\vect{k}_1 + \vect{k}_2\bigr) \, \bar{\rho}^2 \, C_2(1,2) \, \e^{- \frac{\sigma_p^2}{2} \bigl[\bigl(2 \, (1 - \cosmoT_1) \vect{k}_1 + \cosmoT_1 \vect{l}_1\bigr)^2 + \bigl(- 2 \, (1 - \cosmoT_2) \vect{k}_1 + \cosmoT_2 \vect{l}_2\bigr)^2\bigr]} \,,
	\label{eq:cosmo:free_ff_cumulant}
\end{align}
where we neglected contributions due to shot noise by taking the thermodynamic limit, as discussed in \autoref{it:micro:free_cumulants_in_statistically_homogeneous_systems} of \secref{sec:micro:free_cumulants}. This is an excellent approximation on all scales relevant for cosmic structure formation, as the number of dark matter particles contained in any volume of interest is huge.

We also already imposed the constraints on $\vect{k}_2$ and $\vect{l}_2$ set by the delta functions, and defined the short-hand notations
\begin{equation}
	\cosmoT_u \coloneqq \e^{-\frac{1}{2} \eta_u} \,, \quad \cosmoT_{ur} \coloneqq \e^{-\frac{1}{2} (\eta_u - \eta_r)}
\end{equation}
for the time-dependencies introduced by the Green's function. Furthermore, $\sigma_p^2$ denotes the variance of the initial momentum field,
\begin{equation}
	\sigma_p^2 = \frac{1}{3} \fmi{3}{k} \, \frac{P_\delta^\ini(k)}{k^2} \,,
	\label{eq:cosmo:initial_momentum_variance}
\end{equation}
and $C_2$ describes the 2-point correlations of the phase-space density emerging from the free-streaming of particles. Expanding the latter to first order in $P_\delta^\ini$ yields
\begin{equation}
	C_2(1,2) = P_\delta^\ini(k_1) \, \biggl[3 - \biggl(2 - \frac{\vect{k}_1 \cdot \vect{l}_1}{k_1^2}\biggr) \, \cosmoT_1\biggr] \, \biggl[3 - \biggl(2 - \frac{\vect{k}_2 \cdot \vect{l}_2}{k_2^2}\biggr) \, \cosmoT_2\biggr] + \mathcal{O}\Bigl(\bigl(P_\delta^\ini\bigr)^2\Bigr) \,.
	\label{eq:cosmo:undamped_2-particle_contribution_to_free_ff_cumulant}
\end{equation}
As pointed out in \cite{fabis_kinetic_2018}, the full nonlinear expression for $C_2$ given in \eqref{eq:app:2-particle_contribution_to_free_2-point_cumulant} has to be evaluated numerically.

\subsection{Analytic large-scale limit}
\label{sec:cosmo:analytical_large_scale_solution}
To determine the macroscopic propagator for cosmic structure formation, we first insert the expression \eqref{eq:cosmo:free_fF_cumulant} for $\fGfF$ into the functional inverse \eqref{eq:macro:retarded_and_advanced_propagator} defining the causal propagators $\PropR$ and $\PropA$. In \appref{app:computing_the_macroscopic_propagator} we show how this inverse can be computed fully analytically if $\fGfF(1,2)$ evaluated at $\vect{l}_1 = \vect{l}_2 = 0$ is time-translation invariant, i.\,e.~if it only depends on $\eta_1$ and $\eta_2$ in terms of their difference $\eta_{12} \coloneqq \eta_1 - \eta_2$. While this is generally not the case for \eqref{eq:cosmo:free_fF_cumulant}, we can see that it is only the remaining $k_1$-dependent part of the Gaussian damping factor, $\e^{- 2 \sigma_p^2 k_1^2 (\cosmoT_2 - \cosmoT_1)^2}$, which breaks this invariance. For the moment, let us thus consider the large-scale limit $k_1^2 \ll \sigma_p^{-2}$ in which this part of the damping factor becomes negligible and we can follow the steps outlined in \appref{app:computing_the_macroscopic_propagator} to find the analytic solution
\begin{equation}
	\PropR^\largescale(1,2) = \PropA^\largescale(2,1) = \id(1,2) + (2\pi)^3 \dirac\bigl(\vect{k}_1 + \vect{k}_2\bigr) \, (2\pi)^3 \dirac\bigl(\vect{l}_2\bigr) \, \aPropR^\largescale\bigl(\vect{k}_1, \vect{l}_1; \eta_1, \eta_2\bigr)
	\label{eq:cosmo:retarded_and_advanced_propagator_in_large-scale_limit}
\end{equation}
with
\begin{equation}
	\aPropR^\largescale\bigl(\vect{k}_1, \vect{l}_1; \eta_1, \eta_2\bigr) = \frac{3}{5} \, \biggl[\biggl(1 + \frac{\vect{k}_1 \cdot \vect{l}_1}{k_1^2}\biggr) \, \e^{\eta_{12}} - \biggl(1 - \frac{3}{2} \, \frac{\vect{k}_1 \cdot \vect{l}_1}{k_1^2}\biggr) \, \e^{-\frac{3}{2} \eta_{12}}\biggr] \step(\eta_{12}) \, \e^{-\frac{\sigma_p^2}{2} T_1^2 l_1^2} \,.
	\label{eq:cosmo:reduced_retarded_and_advanced_propagator_in_large-scale_limit}
\end{equation}
This immediately fixes the off-diagonal elements of the macroscopic propagator \eqref{eq:macro:explicit_expression_for_the_propagator}.

In this limit, the damping factor in the expression \eqref{eq:cosmo:free_ff_cumulant} for $\fGff$ simplifies as well, and restricting $C_2$ to its linear part becomes a very good approximation, as the nonlinear contributions are expected to be sub-dominant on large scales. Hence, the large-scale limit of $\fGff$ reads
\begin{equation}
	\begin{split}
		G_{ff}^{(0,\text{ls})}(1,2) &= (2\pi)^3 \dirac\bigl(\vect{k}_1 + \vect{k}_2\bigr) \, \bar{\rho}^2 \, P_\delta^\ini(k_1) \\
		&\= \times \biggl[3 - \biggl(2 - \frac{\vect{k}_1 \cdot \vect{l}_1}{k_1^2}\biggr) \, \cosmoT_1\biggr] \, \biggl[3 - \biggl(2 - \frac{\vect{k}_2 \cdot \vect{l}_2}{k_2^2}\biggr) \, \cosmoT_2\biggr] \, \e^{- \frac{\sigma_p^2}{2} \bigl(\cosmoT_1^2 l_1^2 + \cosmoT_2^2 l_2^2\bigr)} \,.
	\end{split}
	\label{eq:cosmo:free_ff_cumulant_in_large-scale_limit}
\end{equation}
Inserting \eqref{eq:cosmo:retarded_and_advanced_propagator_in_large-scale_limit} and \eqref{eq:cosmo:free_ff_cumulant_in_large-scale_limit} into the $ff$-component of \eqref{eq:macro:explicit_expression_for_the_propagator} then yields the remaining statistical propagator,
\begin{equation}
	\begin{split}
		\Propff^\largescale(1,2) &= (2\pi)^3 \dirac\bigl(\vect{k}_1 + \vect{k}_2\bigr) \, \bar{\rho}^2 \, P_\delta^\ini(k_1) \, \e^{\eta_1 + \eta_2} \\
		&\= \times \biggl(1 + \frac{\vect{k}_1 \cdot \vect{l}_1}{k_1^2}\biggr) \, \biggl(1 + \frac{\vect{k}_2 \cdot \vect{l}_2}{k_2^2}\biggr) \, \e^{- \frac{\sigma_p^2}{2} \bigl(\cosmoT_1^2 l_1^2 + \cosmoT_2^2 l_2^2\bigr)} \,,
	\end{split}
	\label{eq:cosmo:ff_component_of_macroscopic_propagator_in_large-scale_limit}
\end{equation}
as shown in detail in \appref{app:computing_the_macroscopic_propagator}.

From \eqref{eq:cosmo:ff_component_of_macroscopic_propagator_in_large-scale_limit} we can infer the large-scale solution of the tree-level density contrast power spectrum,
\begin{align}
	P_\delta^\proplargescale(k_1, \eta_1) &= \left. \frac{1}{\bar{\rho}^2} \fmi{3}{k_2} \Propff^\largescale(1,2) \, \right|_{\substack{\vect{l}_1 = \vect{l}_2 = 0 \\ \eta_2 = \eta_1}} \\
	&= P_\delta^\ini(k_1) \, \e^{2 \eta_1} = P_\delta^\ini(k_1) \, \frac{D_+^2(\eta_1)}{D_+^2(0)} \,,
	\label{eq:cosmo:large-scale_density_contrast_power_spectrum}
\end{align}
using \eqref{eq:micro:spatial_collective_field_cumulants}, \eqref{eq:cosmo:density_contrast_power_spectrum} and \eqref{eq:cosmo:time_coordinate}. We see that we are precisely recovering the well-known linear growth of structures on the largest scales. It can not be stressed enough that at no point in the derivation did we have to employ the Zel'dovich approximation or assume Eulerian fluid dynamics. Instead, this result was derived completely analytically from \emph{Newtonian} particle dynamics alone. To our knowledge, it is the first time this has been achieved, making it one of the main results of our paper.

If we compare \eqref{eq:cosmo:large-scale_density_contrast_power_spectrum} with the large-scale limit of the freely evolved power spectrum,
\begin{align}
	P_\delta^{(0, \text{ls})}(k_1, \eta_1) &= \left. \frac{1}{\bar{\rho}^2} \fmi{3}{k_2} \Gff^{(0, \text{ls})}(1,2) \, \right|_{\substack{\vect{l}_1 = \vect{l}_2 = 0 \\ \eta_2 = \eta_1}} \\
	&= P_\delta^\ini(k_1) \, \bigl(3 - 2 \, \e^{-\frac{1}{2} \eta_1}\bigr)^2 ,
\end{align}
we can see that the growth of the latter is drastically slower and even bounded from above for late times. The reason for this is that, relative to the expanding space-time, freely evolving particles slow down in comoving coordinates since their initial momentum falls behind the cosmic expansion \cite{bartelmann_trajectories_2015}. We conclude that the gravitational particle interactions resummed in the macroscopic propagator, as discussed in \secref{sec:macro:propagator_and_vertices}, precisely compensate the lack of large-scale structure growth caused by this deceleration.

Having the full phase-space information at hand, we can also directly deduce correlations of the momentum density $\vect{\usualpi}$, i.\,e.~the first momentum moment of the phase-space density,
\begin{equation}
	\vect{\usualpi}\bigl(\vect{q}, \eta\bigr) \coloneqq \umi{3}{p} \, \vect{p} \, f\bigl(\vect{q}, \vect{p}, \eta\bigr) \,, \qquad \vect{\usualpi}\bigl(\vect{k}, \eta\bigr) = \left. \ii \frac{\partial}{\partial \vect{l}} \; f\bigl(\vect{k}, \vect{l}, \eta\bigr) \, \right|_{\vect{l}=0} \,.
\end{equation}
Here, we used \eqref{eq:micro:momemtum_moment_operator} to obtain the expression in Fourier space. Its power spectrum is defined via
\begin{equation}
	(2\pi)^3 \dirac\bigl(\vect{k}_1 + \vect{k}_2\bigr) \, P_\usualpi(k_1, \eta) \coloneqq \frac{1}{\bar{\rho}^2} \, \bigl\langle\vect{\usualpi}(\vect{k}_1, \eta) \cdot \vect{\usualpi}(\vect{k}_2, \eta)\bigr\rangle_\C \,,
	\label{eq:cosmo:momentum_density_power_spectrum}
\end{equation}
and has the large-scale limit
\begin{align}
	P_\usualpi^\proplargescale(k_1, \eta_1) &= \left. \frac{1}{\bar{\rho}^2} \fmi{3}{k_2} \, \ii \frac{\partial}{\partial \vect{l}_1} \cdot \ii \frac{\partial}{\partial \vect{l}_2} \, \Propff^\largescale(1,2) \, \right|_{\substack{\vect{l}_1 = \vect{l}_2 = 0 \\ \eta_2 = \eta_1}} \\
	&= \frac{P_\delta^\ini(k_1)}{k_1^2} \, \frac{D_+^2(\eta_1)}{D_+^2(0)} = \frac{P_\delta^\proplargescale(k_1, \eta_1)}{k_1^2} \,,
\end{align}
which agrees with the solution of the linearised Eulerian fluid equations \cite{park_cosmic_2000}.

\subsection{Numerical solution on all scales}
\label{sec:cosmo:numerical_solution_on_al_scales}
Going beyond the large-scale limit requires a numerical evaluation of the macroscopic propagator. We have shown in \appref{app:computing_the_macroscopic_propagator} how this can be reduced to solving a simple matrix equation and computing one- and two-dimensional integrals over time arguments. Performing these steps is in itself computationally cheap and numerically stable.

Although the numerical evaluation of the full nonlinear expression for the free 2-point cumulant $\fGff(1,2)$ entering this computation is generally more challenging, we are able to compute the full tree-level power spectrum of the density contrast,
\begin{equation}
	P_\delta^\prop(k_1, \eta_1) = \left. \frac{1}{\bar{\rho}^2} \fmi{3}{k_2} \Propff(1,2) \, \right|_{\substack{\vect{l}_1 = \vect{l}_2 = 0 \\ \eta_2 = \eta_1}} \,,
	\label{eq:cosmo:full_tree-level_density_contrast_power_spectrum}
\end{equation}
in a numerically fast and stable manner, exploiting that $\fGff(1,2)$ is evaluated at $\vect{l}_1=\vect{l}_2=0$ in this case. Only when computing the full tree-level power spectrum of the momentum density,
\begin{equation}
	P_\usualpi^\prop(k_1, \eta_1) = \left. \frac{1}{\bar{\rho}^2} \fmi{3}{k_2} \, \ii \frac{\partial}{\partial \vect{l}_1} \cdot \ii \frac{\partial}{\partial \vect{l}_2} \, \Propff(1,2) \, \right|_{\substack{\vect{l}_1 = \vect{l}_2 = 0 \\ \eta_2 = \eta_1}} \,,
\end{equation}
which requires to evaluate $\fGff(1,2)$ also for non-vanishing $\vect{l}_1$ and $\vect{l}_2$, our numerical implementation is currently not sufficiently stable. Computing $P_\usualpi^\prop$ will thus be the subject of future work.

\begin{figure}
	\centering
	\subfloat[
		$z=500$, $k_\mathrm{fs} \approx 139 \, h \, \mathrm{Mpc}^{-1}$
		\label{fig:cosmo:normalised_tree-level_powerspectra:z500}]{
		\includegraphics[width=0.48\textwidth]{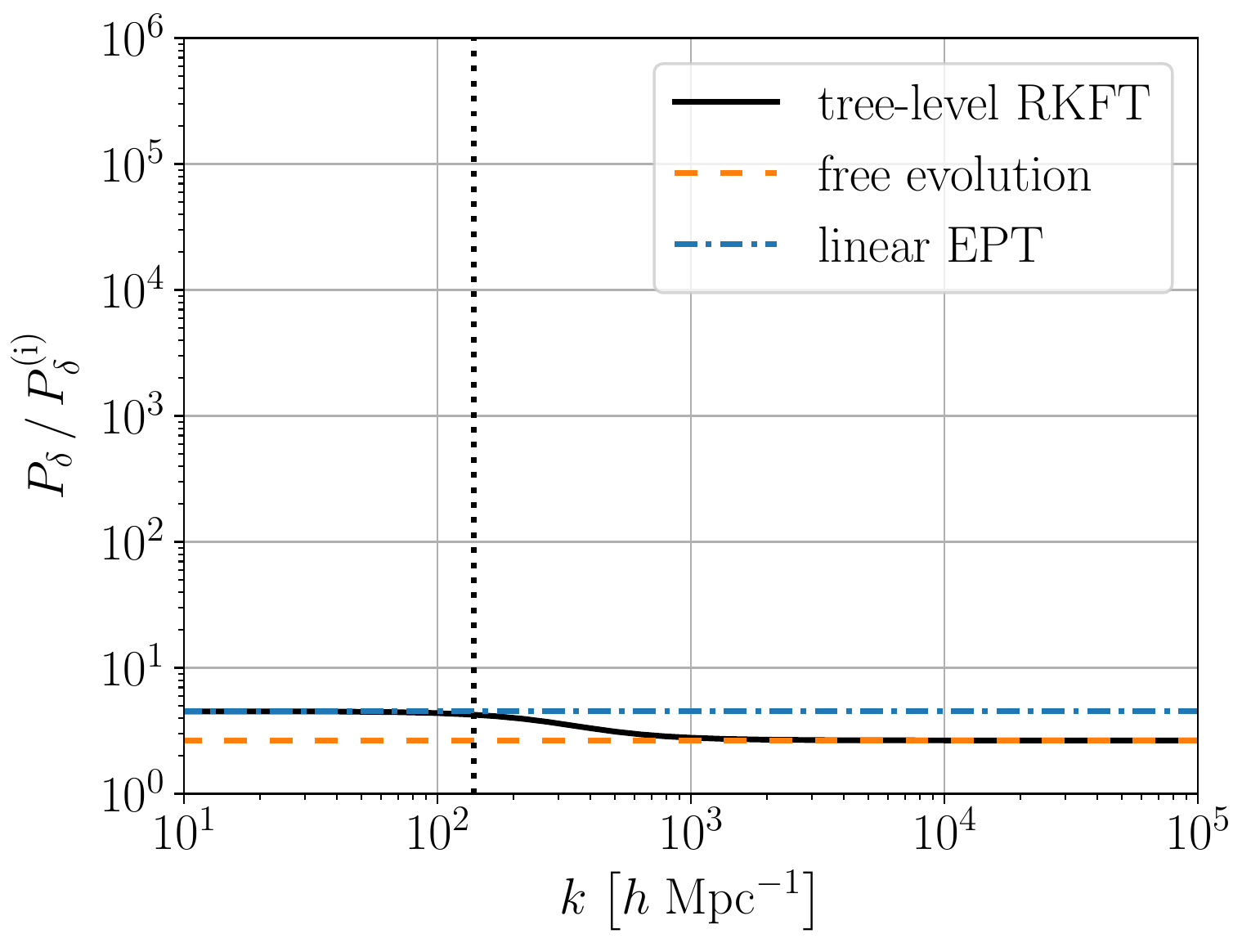}
	}
	\subfloat[
		$z=200$, $k_\mathrm{fs} \approx 79 \, h \, \mathrm{Mpc}^{-1}$
		\label{fig:cosmo:normalised_tree-level_powerspectra:z200}]{
		\includegraphics[width=0.48\textwidth]{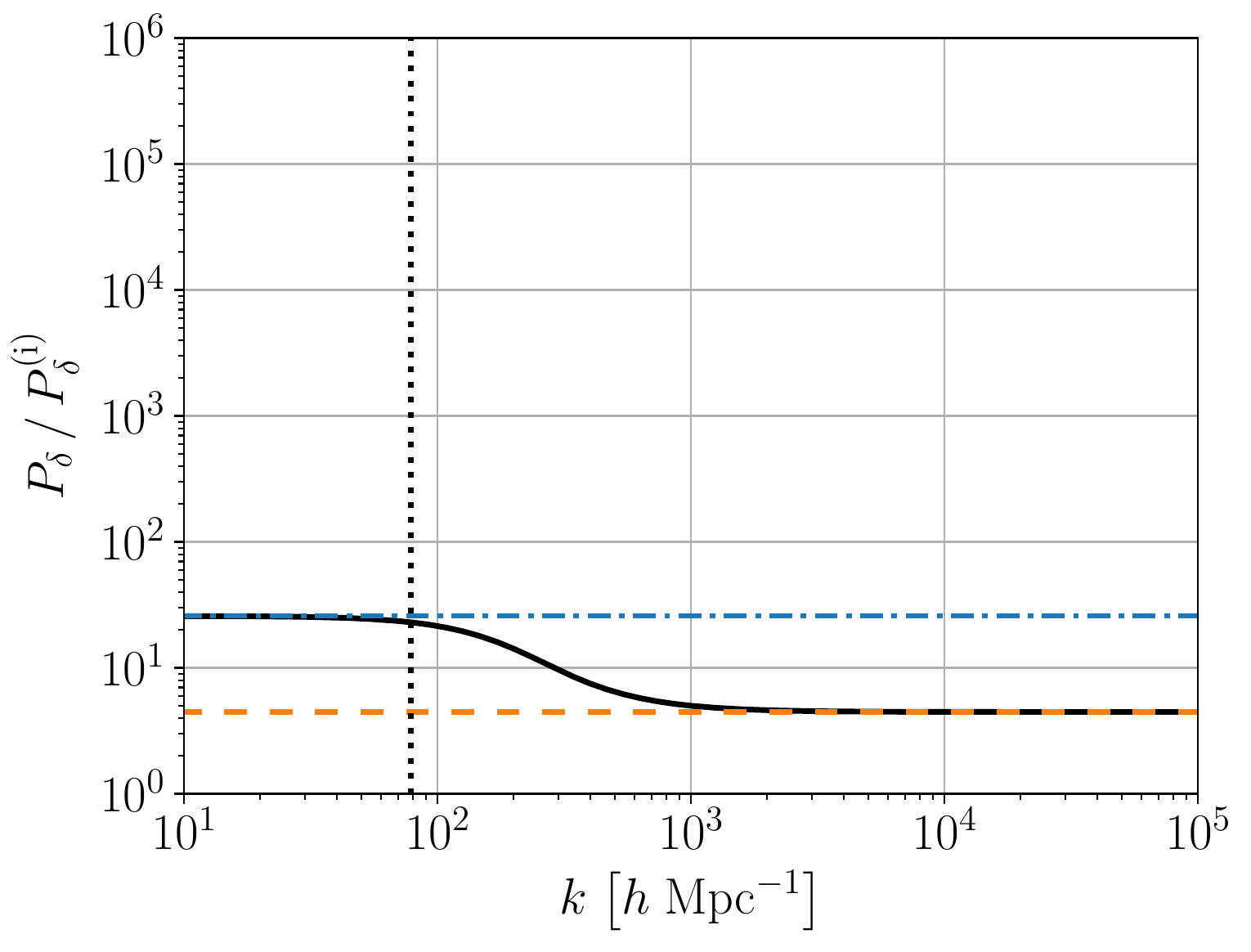}
	}
	\\
	\subfloat[
		$z=20$, $k_\mathrm{fs} \approx 51 \, h \, \mathrm{Mpc}^{-1}$
		\label{fig:cosmo:normalised_tree-level_powerspectra:z20}]{
		\includegraphics[width=0.48\textwidth]{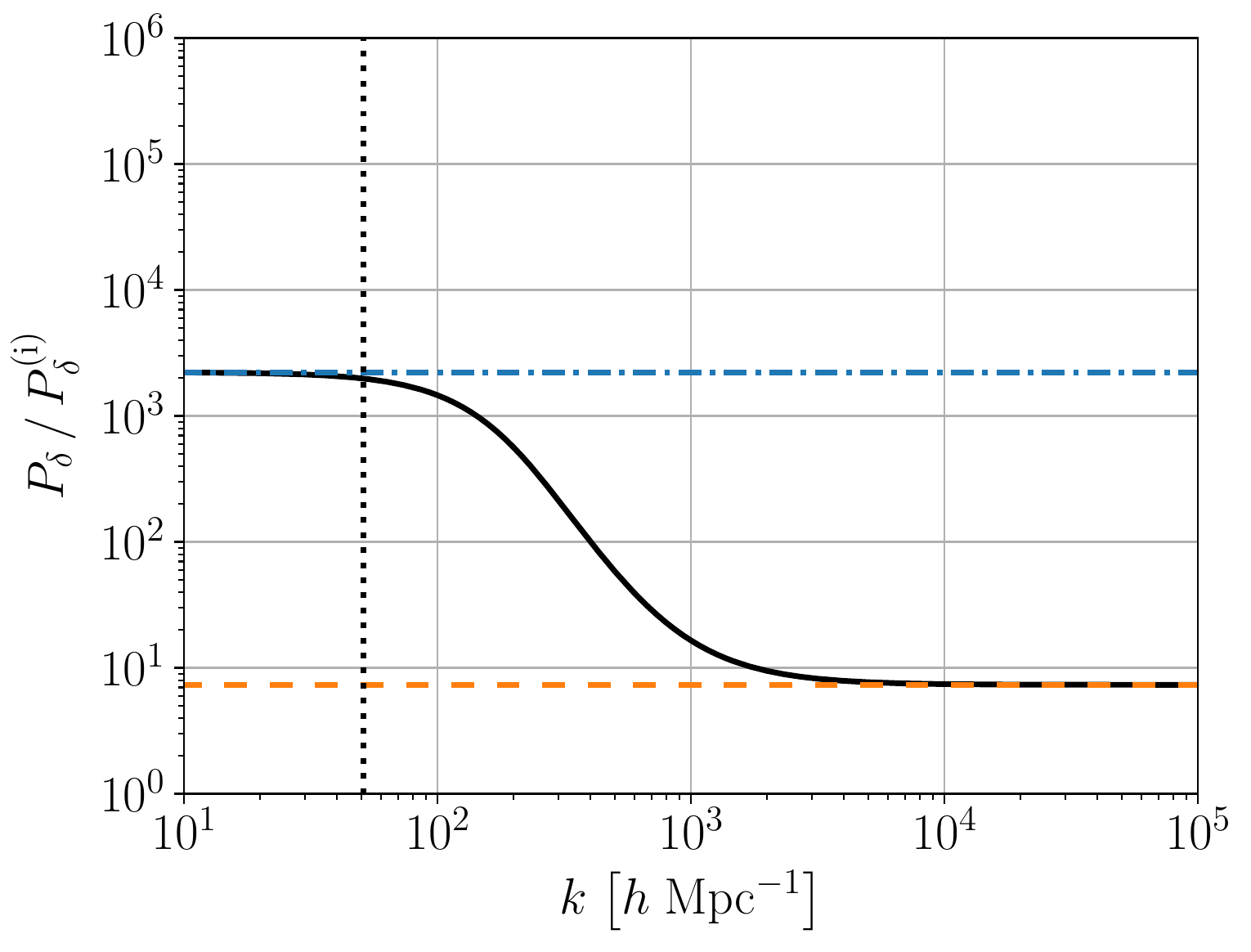}
	}
	\subfloat[
		$z=0$, $k_\mathrm{fs} \approx 45 \, h \, \mathrm{Mpc}^{-1}$
		\label{fig:cosmo:normalised_tree-level_powerspectra:z0}]{
		\includegraphics[width=0.48\textwidth]{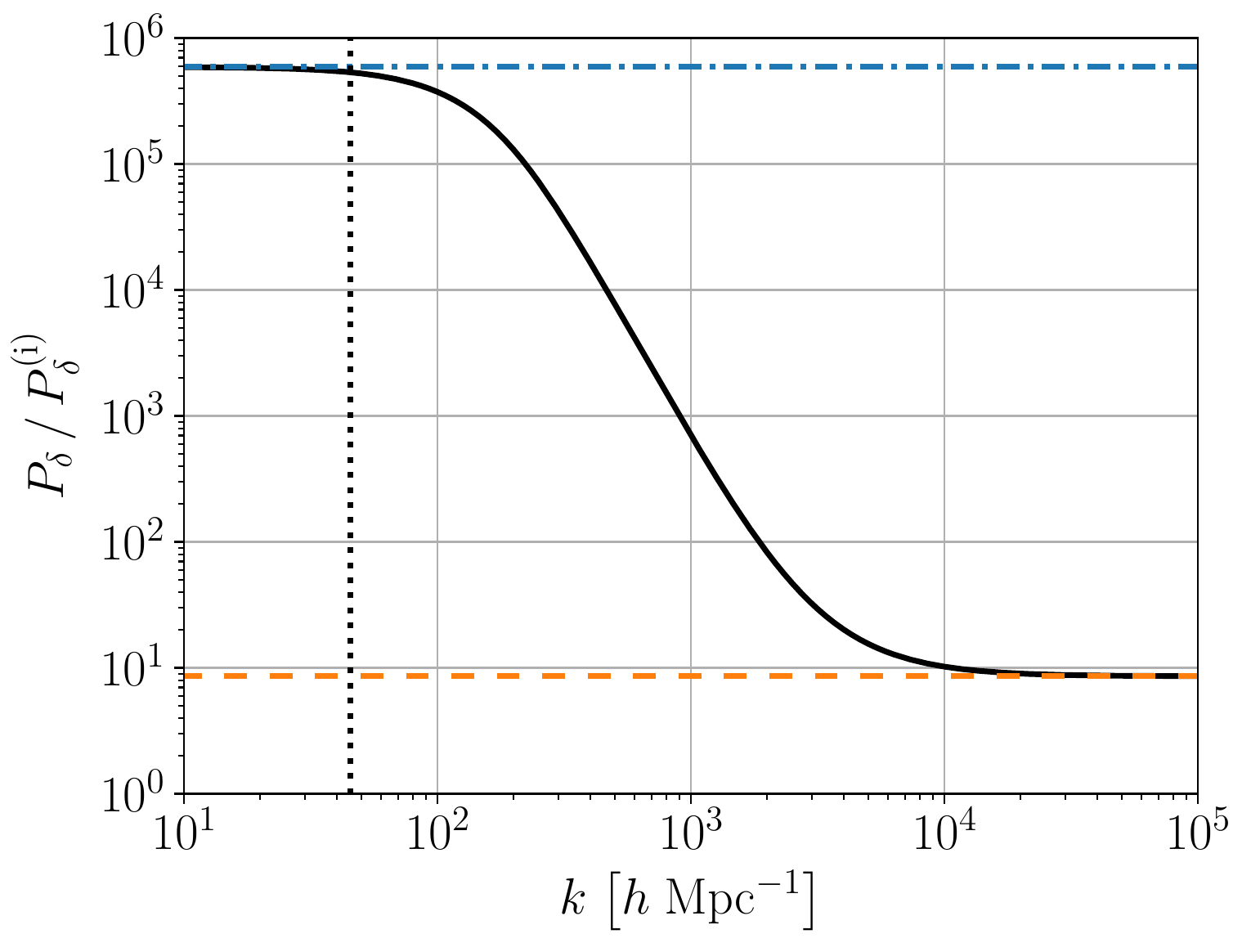}
	}
	\caption{
		Comparison of different density contrast power spectra evolved from the time of CMB decoupling to redshifts 500 (a), 200 (b), 20 (c) and 0 (d): the RKFT tree-level spectrum \eqref{eq:cosmo:full_tree-level_density_contrast_power_spectrum} (black solid), the freely evolved spectrum \eqref{eq:cosmo:freely_evolved_density_contrast_power_spectrum} (orange dashed), and the spectrum obtained from linear Eulerian perturbation theory (EPT), which agrees with \eqref{eq:cosmo:large-scale_density_contrast_power_spectrum} (blue dash-dotted). To reduce the dynamic range, all spectra are divided by the initial spectrum. The RKFT tree-level result follows the linear growth on large scales, drops below it on scales smaller than the particles' mean free-streaming length, and eventually approaches the freely evolved spectrum. The wavenumber $k_\mathrm{fs}$ associated with the mean free-streaming length scale is defined in \eqref{eq:cosmo:mean_free-streaming_wavenumber} and marked with a vertical dotted line for each redshift.
		}
	\label{fig:cosmo:normalised_tree-level_powerspectra}
\end{figure}

For our analysis, we set the initial time to a redshift of $z = 1100$, corresponding approximately to the time of CMB decoupling, and assume the initial power spectrum $P_\delta^\ini$ to be given by a BBKS spectrum \cite{bardeen_statistics_1986} with spectral index $n_\mathrm{s}=1$ normalized such that $\sigma_8=0.8$ today. The resulting full tree-level spectrum $P_\delta^\prop$ is shown in \figref{fig:cosmo:normalised_tree-level_powerspectra} for the four exemplary redshifts 500, 200, 20 and 0. We compare it to the freely evolved spectrum,
\begin{equation}
	P_\delta^\free(k_1, \eta_1) = \left. \frac{1}{\bar{\rho}^2} \fmi{3}{k_2} \fGff(1,2) \, \right|_{\substack{\vect{l}_1 = \vect{l}_2 = 0 \\ \eta_2 = \eta_1}} \,,
	\label{eq:cosmo:freely_evolved_density_contrast_power_spectrum}
\end{equation}
as well as the power spectrum obtained from linear Eulerian perturbation theory (EPT), where the latter corresponds to using our large-scale limit \eqref{eq:cosmo:large-scale_density_contrast_power_spectrum} for all wavenumbers.	To focus on the deviations of the full tree-level result from its large-scale limit, all spectra are divided by the initial spectrum $P_\delta^\ini$ and only wavenumbers $k \geq 10 \, h \, \mathrm{Mpc}^{-1}$ are shown.

While the amplitudes of the spectra of course decrease with redshift, we observe the same qualitative behaviour for all redshifts: The full tree-level result follows the linear EPT spectrum on small wavenumbers, but it drops and eventually approaches the freely evolved spectrum when going to higher wavenumbers. The overall shape of the tree-level spectrum thus interpolates between linear growth on large scales and free evolution of structures on small scales. We further see that the wavenumber where the tree-level spectrum starts to drop below the linear EPT spectrum increases with redshift, while the wavenumber where it starts to follow the freely evolved spectrum decreases.

To understand the origin of this behaviour, we have to take a closer look at the influence of the particles' initial momentum variance $\sigma_p^2$. Since the initial 1-particle momentum distribution function is given by a Maxwell-Boltzmann distribution \cite{fabis_kinetic_2018}, the initial variance $\sigma_p^2$ is related to the mean initial particle momentum via $\bar{p}^\ini = \sqrt{8 / \pi} \, \sigma_p$. Using \eqref{eq:micro:free_trajectories_in_general_form}, the mean distance travelled by a free-streaming particle since the initial time is then obtained by multiplying $\bar{p}^\ini$ with the position-momentum component of the single-particle Green's function,
\begin{equation}
	\Delta \bar{q}^\free(\eta) \coloneqq g_{qp}(\eta, 0) \, \bar{p}^\ini \,.
\end{equation}
Taking the inverse yields an estimate for the wavenumber associated with this mean free-streaming length,
\begin{equation}
	k_\mathrm{fs}(\eta) \coloneqq \frac{1}{\Delta \bar{q}^\free(\eta)} = \frac{\sqrt{\frac{\pi}{32}}}{\bigl(1 - \e^{-\frac{1}{2} \eta}\bigr) \, \sigma_p} \,,
	\label{eq:cosmo:mean_free-streaming_wavenumber}
\end{equation}
where we inserted \eqref{eq:cosmo:single_particle_qp_green's_function}.

For our choice of initial conditions we find $\sigma_p \approx 7.16 \cdot 10^{-3} \, h^{-1} \, \mathrm{Mpc}$, yielding values of $k_\mathrm{fs} \approx 139$, 79, 51 and $45 \, h \, \mathrm{Mpc}^{-1}$ at redshifts $z = 500$, 200, 20 and 0, respectively. We marked the corresponding values by vertical dotted lines in \figref{fig:cosmo:normalised_tree-level_powerspectra}, finding that they match the wavenumbers above which the tree-level power spectrum $P_\delta^\prop$ drops below the linear EPT spectrum very well. This shows that the gravitational interactions resummed by the macroscopic propagator can not fully counteract the dissolution of structures caused by the particles' free-streaming.

The reason for this is that the propagator captures the full nonlinear free-streaming evolution of the phase-space density, but only its linear response to the gravitational interactions. On large scales this reproduces the linear EPT result, as these scales are dominated by contributions linear in the initial power spectrum. On scales smaller than the particles' mean free-streaming length, though, contributions of higher order in the initial spectrum become relevant and suppress the linear growth of structures.

This behaviour is reminiscent of the small-scale suppression of the power spectrum found in Zel'dovich dynamics \cite{bernardeau_large-scale_2002,bartelmann_kinetic_2017} or Renormalized Perturbation Theory (RPT) \cite{crocce_renormalized_2006,crocce_memory_2006}. Unlike those, however, we have seen in \figref{fig:cosmo:normalised_tree-level_powerspectra} that the RKFT tree-level result never drops below the structure growth found for freely evolving particles. This can be explained by the fact that the Neumann series expression \eqref{eq:macro:expansion_of_statistical_propagator_with_powers} of the statistical propagator $\Propff(1,2)$ explicitly contains the free cumulant $\fGff$ as one of its contributions. Thus, if we go to scales smaller than the particles' mean free-streaming length, all the other contributions caused by the gravitational interactions get suppressed until $\fGff$ eventually becomes the dominant contribution. The faster growth of the interacting tree-level contributions compared to the growth of $P_\delta^\free$ further explains why the wavenumber above which $P_\delta^\prop$ follows $P_\delta^\free$ increases towards later times.

Altogether, the full tree-level result for the density contrast power spectrum is found to capture the linear effects introduced by gravitational interactions in a way that is consistent with the underlying free Newtonian particle dynamics. Of course, nonlinear effects of the gravitational interactions will modify the small-scale behaviour found here, as these scales lie far within the nonlinear regime of cosmic structure formation. In upcoming papers we will investigate these effects by including loop corrections.

\section{Summary and conclusions}
\label{sec:conclusions}
Building on our previous work in \cite{bartelmann_microscopic_2016,bartelmann_kinetic_2017,fabis_kinetic_2018}, we developed an exact reformulation of the KFT partition function as a path integral over the macroscopic phase-space density field and a macroscopic auxiliary field only, while preserving all information on the microscopic particle dynamics. This reformulation gave rise to a new macroscopic perturbative expansion that resums an infinite subset of contributions appearing in the original microscopic perturbative series introduced in \cite{bartelmann_microscopic_2016}. We further introduced a diagrammatic language that allows to systematically compute perturbative contributions to cumulants of the phase-space density within this new resummed KFT (RKFT) framework, following a simple set of Feynman rules.

Using this framework to describe the growth of cosmic structures in a standard $\Lambda$CDM cosmology, we calculated the tree-level results for the power spectra of the density contrast and the momentum density of dark matter particles following Newtonian dynamics. In the limit of large scales, these results precisely recovered the well-known linear growth of structures. To our knowledge, this is the first analytic derivation of the emergence of linear large-scale growth from Newtonian particle dynamics alone. At no point in the derivation did we employ the Zel'dovich approximation or assume Eulerian fluid dynamics.

On scales smaller than the particles' mean free-streaming length, the tree-level result for the density contrast power spectrum drops below linear growth, which is reminiscent of the small-scale suppression of the power spectrum found in Zel'dovich dynamics and RPT. But unlike those, our result never drops below the power spectrum resulting from freely streaming particles. This shows that the lowest-order perturbative result within RKFT is able to take into account the linear effects introduced by the gravitational interactions between particles in a way that is consistent with the underlying free Newtonian dynamics.

The small-scale behaviour of the tree-level power spectrum of the momentum density will be investigated in future work once we have implemented the challenging numerical evaluation of the full nonlinear expression for the free 2-point phase-space density cumulant in a sufficiently stable manner. In upcoming papers we will also discuss the computation of loop corrections within the macroscopic perturbation theory and extend the RKFT framework to the treatment of a joint system of dark and baryonic matter.

\acknowledgments{We are grateful for many helpful comments and discussions to Daniel Berg, Daniel Gei\ss, Ivan Kostyuk, Carsten Littek, Robert Reischke and Bj\"orn Malte Sch\"afer. This work was supported in part by the German Excellence Initiative, the Heidelberg Graduate School of Physics and by the Collaborative Research Centre TRR 33 `The Dark Universe' of the German Research Foundation. Most Feynman diagrams were generated using Ti\emph{k}Z-Feynman \cite{ellis_tikz-feynman:_2017}.}

\appendix

\section{General expressions for the free collective-field cumulants}
\label{app:free_cumulants_in_statistically_homogeneous_systems}
In \cite{fabis_kinetic_2018} general expressions for the free collective-field cumulants in statistically homogeneous and isotropic Hamiltonian systems with Gaussian initial conditions have been derived. This means that the particles' initial phase-space coordinates are Poisson sampled from initial density contrast and momentum fields, ${\delta^\ini(\vect{q}) = \bigl(\rho^\ini(\vect{q}) - \bar{\rho}\bigr) / \bar{\rho}}$ and ${\vect{P}^\ini(\vect{q})}$, which together form a Gaussian random field with zero mean and covariance matrix
\begin{align}
	\begin{pmatrix}
		C_{\delta_1 \delta_2} &\vect{C}^\top_{\delta_1 p_2} \\
		\vect{C}_{p_1 \delta_2} &C_{p_1 p_2}
	\end{pmatrix}
	\coloneqq
	\begin{pmatrix}
		\bigl\langle\delta^\ini(\vect{q}_1) \, \delta^\ini(\vect{q}_2)\bigr\rangle &\bigl\langle\delta^\ini(\vect{q}_1) \, \vect{P}^\ini(\vect{q}_2)\bigr\rangle^\top \\
		\bigl\langle\vect{P}^\ini(\vect{q}_1) \, \delta^\ini(\vect{q}_2)\bigr\rangle &\bigl\langle\vect{P}^\ini(\vect{q}_1) \otimes \vect{P}^\ini(\vect{q}_2)\bigr\rangle
	\end{pmatrix} \,.
\end{align}
Homogeneity and isotropy imply that the covariance matrix can only depend on the spatial distance ${|\vect{q}_{12}| \coloneqq |\vect{q}_1 - \vect{q}_2|}$ between two points.

For the non-vanishing 1- and 2-point cumulants used in this work these general expressions read
\begin{align}
	\fGf(1) &= (2\pi)^3 \dirac\bigl(\vect{L}_{q,1}\bigr) \, \bar{\rho} \, \e^{-\frac{\sigma_p^2}{2} \vect{L}_{p,1}^2} \,,
	\label{eq:app:free_f_cumulant_in_statistically_homogeneous_systems} \\
	\begin{split}
		\fGfF(1,2) &= (2\pi)^3 \dirac\bigl(\vect{L}_{q,1} + \vect{L}_{q,2}\bigr) \, (2\pi)^3 \dirac\bigl(\vect{l}_2\bigr) \, \bar{\rho} \, v(k_2, t_2) \\
		&\=\;\times \bigl(-\ii \vect{k}_2\bigr) \cdot \vect{L}_{p,1}(t_2) \, \e^{-\frac{\sigma_p^2}{2} \bigl(\vect{L}_{p,1} + \vect{L}_{p,2}\bigr)^2} \,,
	\end{split}
	\label{eq:app:free_fF_cumulant_in_statistically_homogeneous_systems} \\
	\fGff(1,2) &= (2\pi)^3\dirac\bigl(\vect{L}_{q,1} + \vect{L}_{q,2}\bigr) \biggl[\bar{\rho} \, \e^{-\frac{\sigma_p^2}{2} \bigl(\vect{L}_{p,1} + \vect{L}_{p,2}\bigr)^2} + \bar{\rho}^2 \, C_2(1,2) \, \e^{-\frac{\sigma_p^2}{2} \bigl(\vect{L}_{p,1}^2 + \vect{L}_{p,2}^2\bigr)}\biggr] \,.
	\label{eq:app:free_ff_cumulant_in_statistically_homogeneous_systems}
\end{align}
Here, the vectors $\vect{L}_{q,r}$ and $\vect{L}_{p,r}$ encode the phase shift of the Fourier transformed phase-space density caused by the free particle motion from time $t$ to $t_r$,
\begin{align}
	\vect{L}_{q,r}(t) &\coloneqq \vect{k}_r \, g_{qq}(t_r,t) + \vect{l}_r \, g_{pq}(t_r,t) \,, \qquad \vect{L}_{q,r} \coloneqq \vect{L}_{q,r}(t_\ii) \,,
	\label{eq:app:position_phase_shift_vector} \\
	\vect{L}_{p,r}(t) &\coloneqq \vect{k}_r \, g_{qp}(t_r,t) + \vect{l}_r \, g_{pp}(t_r,t) \,, \qquad \vect{L}_{p,r} \coloneqq \vect{L}_{p,r}(t_\ii) \,.
	\label{eq:app:momentum_phase_shift_vector}
\end{align}
If $t=t_\ii$, we omit writing the time-dependence. Furthermore, $\sigma_p^2$ is the initial 1-point momentum variance,
\begin{equation}
	\sigma_p^2 = \left. \frac{1}{3} \tr{C_{p_1 p_2}} \, \right|_{|\vect{q}_{12}| = 0} \,,
	\label{eq:app:initial_momentum_variance}
\end{equation}
and the function $C_2$ describes the contribution to the 2-point phase-space density cumulant emerging from the correlations between 2 freely evolving particles,
\begin{align}
	C_2(1,2) \coloneqq \umi{3}{q_{12}} \, \e^{-\ii \vect{L}_{q,1} \cdot \vect{q}_{12}} \biggl[\biggl(1 &+ C_{\delta_1 \delta_2} - \ii \vect{L}_{p,1} \cdot \vect{C}_{p_1 \delta_2} - \ii \vect{C}_{\delta_1 p_2} \cdot \vect{L}_{p,2} 
	\label{eq:app:2-particle_contribution_to_free_2-point_cumulant} \\
	&+\bigl(- \ii \vect{L}_{p,1} \cdot \vect{C}_{p_1 \delta_2}\bigr) \bigl(- \ii \vect{C}_{\delta_1 p_2} \cdot \vect{L}_{p,2}\bigr)\biggr) \, \e^{- \vect{L}_{p,1}^\top C_{p_1 p_2} \vect{L}_{p,2}} - 1\biggr] \,.
	\nonumber
\end{align}

In the case of cosmic structure formation, the initial momentum field is irrotational and related to the density contrast field via the continuity equation, which allows us to express all components of the initial covariance matrix in terms of the initial density contrast power spectrum $P_\delta^\ini$ defined in \eqref{eq:cosmo:density_contrast_power_spectrum},
\begin{align}
	C_{\delta_1 \delta_2} &= \fmi{3}{k} \, \e^{\ii \vect{k} \cdot \vect{q}_{12}} \, P_\delta^\ini(k) \,, \\
	\vect{C}_{p_1 \delta_2} = \vect{C}_{\delta_1 p_2} &= \fmi{3}{k} \, \e^{\ii \vect{k} \cdot \vect{q}_{12}} \, P_\delta^\ini(k) \, \frac{\ii \vect{k}}{k^2} \,, \\
	C_{p_1 p_2} &= \fmi{3}{k} \, \e^{\ii \vect{k} \cdot \vect{q}_{12}} \, P_\delta^\ini(k) \, \frac{\vect{k} \otimes \vect{k}}{k^4} \,,	
\end{align}
as was derived in \cite{bartelmann_microscopic_2016}. With these, \eqref{eq:app:initial_momentum_variance} and \eqref{eq:app:2-particle_contribution_to_free_2-point_cumulant} become
\begin{equation}
	\sigma_p^2 = \frac{1}{3} \fmi{3}{k} \, \frac{P_\delta^\ini(k)}{k^2}
\end{equation}
and
\begin{equation}
	C_2(1,2) = P_\delta^\ini\bigl(\vect{L}_{q,1}\bigr) \, \biggl(1 + \frac{\vect{k}_1 \cdot \vect{L}_{p,1}}{k_1^2}\biggr) \, \biggl(1 + \frac{\vect{k}_2 \cdot \vect{L}_{p,2}}{k_2^2}\biggr) + \mathcal{O}\Bigl(\bigl({P_\delta^\ini}\bigr)^2\Bigr) \,,
\end{equation}
where we expanded $C_2$ only to first order in $P_\delta^\ini$. Evaluating the full nonlinear expression for $C_2$ requires to perform the Fourier transform in \eqref{eq:app:2-particle_contribution_to_free_2-point_cumulant} numerically.

\section{Computation of the macroscopic propagator}
\label{app:computing_the_macroscopic_propagator}
The causal macroscopic propagators $\PropR$ and $\PropA$ are given by the functional inverse \eqref{eq:macro:retarded_and_advanced_propagator} which is defined as the solution of the integral equation
\begin{equation}
	\usi{\bar{1}} \Bigl(\id(1, \bar{1}) - \ii \fGfF(1, \bar{1})\Bigr) \; \PropR(-\bar{1}, 2) = \id(1,2) \,,
	\label{eq:app:integral_equation_for_full_retarded_propagator}
\end{equation}
with the identity 2-point function $\id$ introduced in \eqref{eq:macro:identity_2-point_function}. The physical systems we are most interested in are statistically homogeneous and have a free Hamiltonian that only depends on the particle momenta but not their positions. The latter implies $g_{qq}(t,t') = \step(t-t')$ and $g_{pq}(t,t') = 0$, as shown in \cite{bartelmann_microscopic_2016}. In this case we can conclude from \eqref{eq:app:free_fF_cumulant_in_statistically_homogeneous_systems} and \eqref{eq:app:position_phase_shift_vector} that $\fGfF$ can be written as
\begin{equation}
	\fGfF(1,2) = (2\pi)^3 \dirac\bigl(\vect{k}_1 + \vect{k}_2\bigr) \, (2\pi)^3 \dirac\bigl(\vect{l}_2\bigr) \, \afGfF\bigl(\vect{k}_1, \vect{l}_1; t_1, t_2\bigr) \propto \step(t_1 - t_2) \,,
	\label{eq:app:separating_operator_structure_from_free_fF_cumulant}
\end{equation}
where we introduced the reduced cumulant $\afGfF$ which exploits the constraints set by the delta functions,
\begin{equation}
	\afGfF\bigl(\vect{k}_1, \vect{l}_1; t_1, t_2\bigr) \coloneqq \fmi{6}{s_2} \fGfF(1,2) \,.
	\label{eq:app:reduced_free_fF_cumulant}
\end{equation}
Inserting \eqref{eq:app:separating_operator_structure_from_free_fF_cumulant} into the Neumann series \eqref{eq:macro:expansion_of_retarded_and_advanced_propagator_with_integrals} then suggests the following ansatz for $\PropR$,
\begin{equation}
	\PropR(1,2) = \id(1,2) + (2\pi)^3 \dirac\bigl(\vect{k}_1 + \vect{k}_2\bigr) \, (2\pi)^3 \dirac\bigl(\vect{l}_2\bigr) \, \aPropR\bigl(\vect{k}_1, \vect{l}_1; t_1, t_2\bigr) \,,
	\label{eq:app:reduced_retarded_propagator}
\end{equation}
where $\aPropR\bigl(\vect{k}_1, \vect{l}_1; t_1 t_2\bigr) \propto \step(t_1 - t_2)$. This reduces \eqref{eq:app:integral_equation_for_full_retarded_propagator} to the following integral equation for $\aPropR$,
\begin{equation}
	\aPropR\bigl(\vect{k}_1, \vect{l}_1; t_1, t_2\bigr) = \ii \afGfF\bigl(\vect{k}_1, \vect{l}_1; t_1, t_2\bigr) + \bsi{t_{\bar{1}}}{t_2}{t_1} \,  \ii\afGfF\bigl(\vect{k}_1, \vect{l}_1; t_1, t_{\bar{1}}\bigr) \, \aPropR\bigl(\vect{k}_1, 0; t_{\bar{1}}, t_2\bigr) \,,
	\label{eq:app:integral_equation_for_reduced_retarded_propagator}
\end{equation}
which can be solved in two steps. First, we solve it in the case $\vect{l}_1 = 0$,
\begin{equation}
	\aPropR\bigl(\vect{k}_1, 0; t_1, t_2\bigr) = \ii \afGfF\bigl(\vect{k}_1, 0; t_1, t_2\bigr) + \bsi{t_{\bar{1}}}{t_2}{t_1} \, \ii\afGfF\bigl(\vect{k}_1, 0; t_1, t_{\bar{1}}\bigr) \, \aPropR\bigl(\vect{k}_1, 0; t_{\bar{1}}, t_2\bigr) \,,
	\label{eq:app:integral_equation_for_reduced_retarded_propagator_at_zero_l}
\end{equation}
which can be done independently for all possible values of $\vect{k}_1$. Afterwards we insert the result $\aPropR\bigl(\vect{k}_1, 0; t_1, t_2\bigr)$ of this into the right-hand side of \eqref{eq:app:integral_equation_for_reduced_retarded_propagator} and perform the time integral to obtain the full $\vect{l}_1$-dependent solution $\aPropR\bigl(\vect{k}_1, \vect{l}_1; t_1, t_2\bigr)$.

If $\afGfF\bigl(\vect{k}_1, 0; t_1, t_2\bigr)$ depends on $t_1$ and $t_2$ only in terms of their difference $t_{12} \coloneqq t_1 - t_2$, then according to \eqref{eq:macro:expansion_of_retarded_and_advanced_propagator_with_integrals} the same should hold for $\aPropR\bigl(\vect{k}_1, 0; t_1, t_2\bigr)$, and the time integral in \eqref{eq:app:integral_equation_for_reduced_retarded_propagator_at_zero_l} becomes a convolution,
\begin{equation}
	\aPropR\bigl(\vect{k}_1, 0; t_{12}, 0\bigr) = \ii \afGfF\bigl(\vect{k}_1, 0; t_{12}, 0\bigr) + \bsi{t_{\bar{1}2}}{0}{t_{12}} \, \ii\afGfF\bigl(\vect{k}_1, 0; t_{12} - t_{\bar{1}2}, 0\bigr) \, \aPropR\bigl(\vect{k}_1, 0; t_{\bar{1}2}, 0\bigr) \,,
\end{equation}
with $t_{\bar{1}2} \coloneqq t_{\bar{1}} - t_2$. In this case we can turn the integral equation into an algebraic equation by means of a Laplace transform with respect to $t_{12}$,
\begin{equation}
	\begin{split}
		\L_{t_{12}}\bigl[\aPropR\bigl(\vect{k}_1, 0; t_{12}, 0\bigr)\bigr](z) = \+ &\L_{t_{12}}\bigl[\ii \afGfF\bigl(\vect{k}_1, 0; t_{12}, 0\bigr)\bigr](z) \\
		+ &\L_{t_{12}}\bigl[\ii \afGfF\bigl(\vect{k}_1, 0; t_{12}, 0\bigr)\Bigr](z) \; \L_{t_{12}}\bigl[\aPropR\bigl(\vect{k}_1, 0; t_{12}, 0\bigr)\bigr](z) \,,
	\end{split}
\end{equation}
where $z$ denotes the complex frequency conjugate to $t_{12}$. Bringing $\aPropR$ to one side of the equation and performing an inverse Laplace transform then yields the solution of \eqref{eq:app:integral_equation_for_reduced_retarded_propagator_at_zero_l},
\begin{equation}
   \aPropR\bigl(\vect{k}_1, 0; t_{12}, 0\bigr) = \L^{-1}_z\left[\frac{\L_{t_{12}}\bigl[\ii \afGfF\bigl(\vect{k}_1, 0; t_{12}, 0\bigr)\bigr]}{1 - \L_{t_{12}}\bigl[\ii \afGfF\bigl(\vect{k}_1, 0; t_{12}, 0\bigr)\bigr]}\right](t_{12}) \,.
   \label{eq:app:reduced_retarded_propagator_via_inverse_Laplace_transform}
\end{equation}

Generally, however, the time-dependence of $\afGfF$ will not be that simple and we have to determine $\aPropR$ numerically. This can be done by discretizing the time arguments in \eqref{eq:app:integral_equation_for_reduced_retarded_propagator_at_zero_l} and solving the resulting matrix equation for each $\vect{k}_1$-value of interest. Note that, due to their retarded causal structure, $\afGfF$ and $\aPropR$ become lower triangular matrices after discretization. Hence, this matrix equation can be solved with minimal computational effort via forward substitution.

Once the causal propagators are known we can insert them into the relation \eqref{eq:macro:explicit_expression_for_the_propagator} for the complete macroscopic propagator, which immediately fixes the off-diagonal components. The computation of the remaining statistical propagator then reduces to performing the following time integrals,
\begin{align}
	\Propff(1,2) &= \usi{\bar{1}} \usi{\bar{2}} \, \PropR(1, \bar{1}) \, \fGff(-\bar{1}, \bar{2}) \, \PropA(-\bar{2}, 2) \\
	\begin{split}
		&= (2\pi)^3 \dirac\bigl(\vect{k}_1 + \vect{k}_2\bigr) \, \biggl[\afGff\bigl(\vect{k}_1, \vect{l}_1, \vect{l}_2; t_1, t_2\bigr) \\
		&\= + \bsi{t_{\bar{1}}}{t_\ii}{t_1} \, \aPropR\bigl(\vect{k}_1, \vect{l}_1; t_1, t_{\bar{1}}\bigr) \, \afGff\bigl(\vect{k}_1, 0, \vect{l}_2; t_{\bar{1}}, t_2\bigr) \\
		&\= + \bsi{t_{\bar{2}}}{t_\ii}{t_2} \, \afGff\bigl(\vect{k}_1, \vect{l}_1, 0; t_1, t_{\bar{2}}\bigr) \, \aPropA\bigl(\vect{k}_1, \vect{l}_2; t_{\bar{2}}, t_2\bigr) \\
		&\= + \bsi{t_{\bar{1}}}{t_\ii}{t_1} \bsi{t_{\bar{2}}}{t_\ii}{t_2} \, \aPropR\bigl(\vect{k}_1, \vect{l}_1; t_1, t_{\bar{1}}\bigr) \, \afGff\bigl(\vect{k}_1, 0, 0; t_{\bar{1}}, t_{\bar{2}}\bigr) \, \aPropA\bigl(\vect{k}_1, \vect{l}_2; t_{\bar{2}}, t_2\bigr)\biggr] \,,
	\end{split}
	\label{eq:app:remaining_time_integrals_in_computation_of_ff_component_of_macroscopic_propagator}
\end{align}
where we defined $\aPropA$ analogously to \eqref{eq:app:reduced_retarded_propagator} and $\afGff$ as
\begin{equation}
	\afGff\bigl(\vect{k}_1, \vect{l}_1, \vect{l}_2; t_1, t_2\bigr) \coloneqq \fmi{3}{k_2} \fGff(1,2) \,.
	\label{eq:app:reduced_free_ff_cumulant}
\end{equation}
Depending on whether there exist analytic expressions for $\aPropR$, $\aPropA$ and $\afGff$, these integrals can either be performed analytically or have to be evaluated numerically.

In the case of cosmic structure formation in the large-scale limit, as discussed in \secref{sec:cosmo:analytical_large_scale_solution}, we find
\begin{equation}
	\ii \tilde{G}_{f\F}^{(0,\text{ls})}\bigl(\vect{k}_1, \vect{l}_1; \eta_1, \eta_2\bigr) = \frac{3}{2} \, \biggl[2 - \biggl(2 - \frac{\vect{k}_1 \cdot \vect{l}_1}{k_1^2}\biggl) \, \e^{-\frac{1}{2} \eta_{12}}\biggr] \, \step(\eta_{12}) \, \e^{- \frac{\sigma_p^2}{2} \cosmoT_1^2 l_1^2} \,.
	\label{eq:app:reduced_free_fF_cumulant_for large-scale_cosmic_structure_formation}
\end{equation}
Evaluating this at $\vect{l}_1 = 0$ and Laplace transforming it yields
\begin{equation}
	\L_{\eta_{12}}\Bigl[\ii \tilde{G}_{f\F}^{(0,\text{ls})}\bigl(\vect{k}_1, 0; \eta_{12}, 0\bigr)\Bigr](z) = \frac{3}{z} - \frac{3}{z+\frac{1}{2}} \,,
\end{equation}
which we can insert into \eqref{eq:app:reduced_retarded_propagator_via_inverse_Laplace_transform} to find
\begin{equation}
	\aPropR^\largescale\bigl(\vect{k}_1, 0; \eta_{12}, 0\bigr) = \L^{-1}_z\biggl[\frac{3}{2 \, z^2 + z - 3}\biggr](\eta_{12}) = \frac{3}{5} \, \Bigl(\e^{\eta_{12}} - \e^{-\frac{3}{2} \eta_{12}}\Bigr) \,.
\end{equation}
Inserting this result as well as \eqref{eq:app:reduced_free_fF_cumulant_for large-scale_cosmic_structure_formation} into the right-hand side of \eqref{eq:app:integral_equation_for_reduced_retarded_propagator} and performing the time integral leads to the expression \eqref{eq:cosmo:retarded_and_advanced_propagator_in_large-scale_limit} for the causal propagators. Using that expression together with  the large-scale limit \eqref{eq:cosmo:free_ff_cumulant_in_large-scale_limit} of $\fGff$ in \eqref{eq:app:remaining_time_integrals_in_computation_of_ff_component_of_macroscopic_propagator}, we finally find the expression \eqref{eq:cosmo:ff_component_of_macroscopic_propagator_in_large-scale_limit} for the statistical propagator after performing the remaining time integrals.

\section{Proofs of the Feynman rules}
\label{app:proofs_of_the_Feynman_rules}

\subsection{Causality rule}
\label{app:proofs_of_the_Feynman_rules:causality_rule}
\begin{figure}
	\centering
	\raisebox{-2ex}{\includegraphics{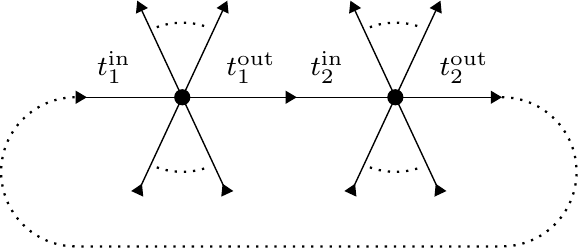}}
	\caption{Diagram with a closed loop in its time flow.}
	\label{fig:app:closed_time-flow_loop}
\end{figure}

Consider a diagram which has only incoming arrows on its outer legs or contains a subdiagram which does so. From the continuity of the time-flow and the fact that there exist no propagators or vertices with only incoming arrows, $\Propbb = 0$ and $\Vert_{f \dotsm f} = 0$, it then follows that there has to be at least one vertex in this diagram from which every possible path along the time-flow ends up in
a closed loop, as illustrated in \figref{fig:app:closed_time-flow_loop}.

In general, there might be multiple possible loops, as the involved vertices might have multiple outgoing legs. In this case we consider the loop that is obtained by choosing for each vertex the path through the outgoing leg evaluated at the latest time argument. Then, the causal structures of the propagators \eqref{eq:macro:causal_structure_of_retarded_and_advanced_propagator} and vertices \eqref{eq:macro:causal_structure_of_vertex} imply that the diagram can only be non-vanishing if the time arguments along the loop satisfy $t_1^{\text{in}} \leq t_1^{\text{out}} \leq t_2^{\text{in}} \leq t_2^{\text{out}} \leq \dotsm \leq t_1^{\text{in}}$.

A closer inspection allows to tighten this restriction even further, though. For this, we first insert the proportionality relation \eqref{eq:micro:potential_gradients_and_delta_functions_of_l_in_free_cumulants} of the free collective-field cumulants into the expressions \eqref{eq:macro:vertices} and \eqref{eq:macro:expansion_of_retarded_and_advanced_propagator_with_powers} for the vertices and the causal propagators, to infer
\begin{align}
	\Vert_{\beta \dotsm \beta f \dotsm f}(1,\dotsc,n_\beta,1',\dotsc,n'_f) &\propto \dirac\bigl(\vect{l}_{r'}\bigr) \quad \forall \; r' \in \{1',\dotsc,n'_f\} \,,
	\label{eq:app:delta_functions_of_l_in_vertices} \\
	\Propfb(1,2) = \Propbf(2,1) =  -\ii \PropR(1,2) &=  - \ii \underbrace{\id(1,2) \vphantom{\sum_{n=1}^\infty}}_{\mathclap{\propto \dirac\bigl(\vect{l}_1 + \vect{l}_2\bigr)}} - \ii \underbrace{\sum_{n=1}^\infty \, \Bigl(\ii \fGfF\Bigr)^n(1,2)}_{\propto \dirac\bigl(\vect{l}_2\bigr)} \,.
	\label{eq:app:delta_functions_of_l_in_off-diagonal_propagator}
\end{align}
Combining both of these relations then yields
\begin{equation}
	\raisebox{-0.5\height+0.2ex}{\includegraphics{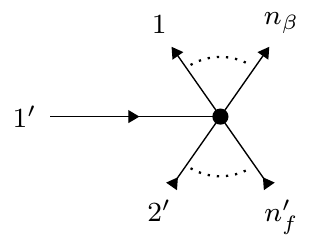}} = \usi{\bar{1}} \, \Propbf(1',\bar{1}) \, \Vert_{f f \dotsm f \beta \dotsm \beta}(-\bar{1},2',\dotsc,n'_f,1,\dotsc,n_\beta) \, \propto \dirac\bigl(\vect{l}_{1'}\bigr) \,.
	\label{eq:app:dirac_function_of_l_in_loop_segment}
\end{equation}
Using this result in \figref{fig:app:closed_time-flow_loop} tells us that the $\vect{l}$-argument at every outgoing $\beta$-leg of a vertex through which the loop passes is set to zero, which means that we actually have to apply the stricter case of the vertices' causal restriction \eqref{eq:macro:causal_structure_of_vertex}. As a consequence, the considered type of diagram could in fact only be non-vanishing if the time arguments along the loop satisfied $t_1^{\text{in}} < t_1^{\text{out}} \leq t_2^{\text{in}} < t_2^{\text{out}} \leq \dotsm \leq t_1^{\text{in}}$. This, however, is always a contradiction and thus the diagram vanishes identically, concluding the proof of the \hyperref[it:macro:causality_rule]{Causality rule}.

\subsection{Homogeneity rule}
\label{app:proofs_of_the_Feynman_rules:homogeneity_rule}
Let the system under consideration be statistically homogeneous, and consider a diagram that contains a so-called tadpole subdiagram, i.\,e.~a subdiagram which is connected to the rest of the diagram solely via a single propagator. We first use that due to the conservation of spatial Fourier modes, \eqref{eq:macro:conservation_of_spatial_fourier_modes_in_propagator} and \eqref{eq:macro:conservation_of_spatial_fourier_modes_in_vertices}, such a tadpole subdiagram satisfies
\begin{equation}
	\raisebox{-0.5\height+0.6ex}{\includegraphics{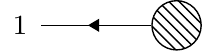}}\propto \dirac\bigl(\vect{k}_1\bigr) \,,
	\label{eq:app:tadpole_diagram}
\end{equation}
where the hatched circle represents a subdiagram with no further outer legs than the one attached to the propagator on the left. Note that we do not have to consider tadpole diagrams using the other two possible propagators since they would vanish identically according to the \hyperref[it:macro:causality_rule]{Causality rule}.

Then, we insert the proportionality relation \eqref{eq:micro:potential_gradients_and_delta_functions_of_l_in_free_cumulants} of the free collective-field cumulants into the definition \eqref{eq:macro:vertices} of the vertices, to obtain
\begin{equation}
	\Vert_{\beta \dotsm \beta f \dotsm f}(1,\dotsc,n_\beta,1',\dotsc,n'_f) \propto \vect{k}_{r'} \, v(k_{r'}, t_{r'}) \quad \forall \; r' \in \{1',\dotsc,n'_f\} \,.
	\label{eq:app:potential_gradients_in_vertices}
\end{equation}
If we now attach the tadpole diagram to the $f$-leg of some vertex, \eqref{eq:app:tadpole_diagram} and \eqref{eq:app:potential_gradients_in_vertices} together imply that the resulting diagram vanishes identically if $\vect{k} \, v(k,t) = 0$ at $\vect{k} = 0$. Fourier transforming this condition to real space,
\begin{equation}
	0 \stackrel{!}{=} \left. \vect{k} \, v(k,t) \, \right|_{\vect{k} = 0} = - \ii \umi{3}{q} \, \vect{\nabla}_q v(q,t) \,,
	\label{eq:app:condition_for_vanishing_tadpole_contribution}
\end{equation}
reveals that it is always satisfied, as the integrand on the right hand side switches sign under a sign flip of $\vect{q}$. Physically, this corresponds to the fact that the volume integral of a particle's central force field over the whole space vanishes. This concludes the proof of the \hyperref[it:macro:homogeneity_rule]{Homogeneity rule}.

Note that for the gravitational potential \eqref{eq:cosmo:single_particle_potential} the condition \eqref{eq:app:condition_for_vanishing_tadpole_contribution} apparently does not hold in Fourier space, even though it is always satisfied in real space. The reason for this is that the Fourier transform of an $1/q$ potential is actually only properly defined as the limit of a large-scale regularisation,
\begin{equation}
	v(k) \propto \lim_{k_\C \rightarrow 0} \umi{3}{q} \, \frac{\e^{-k_\C q}}{q} \, \e^{-\ii \vect{k} \cdot \vect{q}} \propto \lim_{k_\C \rightarrow 0} \frac{1}{k^2 + k_\C^2} \,.
	\label{eq:app:proper_definition_of_gravitational_potential}
\end{equation}
Using this proper definition, the condition \eqref{eq:app:condition_for_vanishing_tadpole_contribution} is indeed satisfied for the Fourier transformed gravitational potential. This is related to the so-called ``Jeans swindle" \cite{valageas_new_2004}. In practice, however, one rarely encounters situations where using \eqref{eq:app:proper_definition_of_gravitational_potential} makes a difference. In particular, this is true for all calculations in \secref{sec:cosmo}, which is why we directly set $k_\C$ to zero in \eqref{eq:cosmo:single_particle_potential}. Nevertheless, one should always keep \eqref{eq:app:proper_definition_of_gravitational_potential} in mind.

\section{Equations of motion for point particles in an expanding space-time}
\label{app:equations_of_motion_for_a_dm_particle}
The Lagrangian of a single classical point particle of mass $m$ in an expanding space-time, expressed in comoving coordinates $\vect{q} = \vect{r} / a$ and cosmic time $t$, is given by
\begin{equation}
	L\bigl(\vect{q}, \dot{\vect{q}}, t\bigr) = \frac{m}{2} \, a^2 \dot{\vect{q}}^2 - m \, V(\vect{q}, t) \,,
\end{equation}
with the Newtonian
gravitational potential $V$ satisfying the Poisson equation
\begin{equation}
	\vect{\nabla}_q^2 V = \frac{4 \pi G}{a} \, (\rho_\m - \bar{\rho}_\m) \,,
	\label{eq:app:normal_poisson_equation}
\end{equation}
as derived in \cite{bartelmann_trajectories_2015,peebles_large-scale_1980}.
Here, $\rho_\m$ denotes the comoving mass density of the total cosmic matter content, and $\bar{\rho}_\m$ is its mean value, which is constant in time.

Following the steps detailed in \cite{bartelmann_trajectories_2015}, we transform the Lagrangian to the new time coordinate $\eta(t) \coloneqq \log{\bigl(D_+(t) / D_+(t_\ii)\bigr)}$ and deduce the resulting Hamiltonian equations of motion,
\begin{align}
	\td{\vect{q}}{\eta} &= \frac{\vect{p}_\text{can}}{m a^2 H f_+} \,, 
	\label{eq:app:Hamiltonian_EOM_of_DM_particle_position} \\
	\td{\vect{p}_\text{can}}{\eta} &= - \frac{m}{H f_+} \, \vect{\nabla}_q V \,,
	\label{eq:app:Hamiltonian_EOM_of_DM_particle_momentum}
\end{align}
with the canonically conjugate momentum $\vect{p}_\text{can}$, the growth function $f_+ \coloneqq \d \ln{D_+} / \d \ln{a}$ and the Hubble parameter $H \coloneqq \dot{a} / a$.

For our purposes, though, it is more convenient to work with the rescaled momentum variable
\begin{equation}
	\vect{p} \coloneqq \frac{\vect{p}_{\text{can}}}{m a^2 H f_+}
\end{equation}
instead, since this choice leads to the much simpler equations of motion \eqref{eq:cosmo:EOM_of_DM_particle_position} and \eqref{eq:cosmo:EOM_of_DM_particle_momentum}.\footnote{Note that our assumption of the Jacobian determinant of the equations of motion being constant, as discussed below \eqref{eq:micro:transforming_delta_function_in_classical_trajectory_to_delta_function_in_EOM}, still holds for \eqref{eq:cosmo:EOM_of_DM_particle_position} and \eqref{eq:cosmo:EOM_of_DM_particle_momentum}, even though $(\vect{q}, \vect{p}_{\text{can}}) \rightarrow (\vect{q}, \vect{p})$ is no canonical transformation. In fact, any invertible time-dependent linear transformation of $(\vect{q}, \vect{p}_{\text{can}})$ preserves this property, as can be shown by a slight generalisation of the argument used in \cite{gozzi_hidden_1989}.} To obtain these, we expressed $\vect{p}_{\text{can}}$ in \eqref{eq:app:Hamiltonian_EOM_of_DM_particle_position} and \eqref{eq:app:Hamiltonian_EOM_of_DM_particle_momentum} in terms of $\vect{p}$, defined the rescaled gravitational potential
\begin{equation}
	\tilde{V} \coloneqq \frac{V}{a^2 f_+^2 H^2} \,,
\end{equation}
and used $H f_+ = \dot{D}_+ / D_+$ as well as the fact that $D_+$ solves the linearised density perturbation evolution equation
\begin{equation}
	\ddot{D}_+ + 2 H \dot{D}_+ = \frac{3}{2} \, \Omega_\m H^2 D_+ \,,
\end{equation}
see \cite{bernardeau_large-scale_2002}. Here,
\begin{equation}
	\Omega_\m = \frac{8 \pi G}{3 a^3 H^2} \, \bar{\rho}_\m
	\label{eq:app:dimensionless_matter_density_parameter}
\end{equation}
is the dimensionless matter density parameter. The rescaled potential $\tilde{V}$ satisfies the modified Poisson equation \eqref{eq:cosmo:poisson_equation}, where we additionally assumed the whole matter content to be made up of point particles of mass $m$, such that $\rho_\m = m \Phi_\rho$.

\bibliography{Bibliography}

\providecommand{\href}[2]{#2}\begingroup\raggedright\begin{thebibliography}{10}

\bibitem{bernardeau_large-scale_2002}
F.~Bernardeau, S.~Colombi, E.~Gazta{\~n}aga and R.~Scoccimarro,
  \emph{Large-scale structure of the universe and cosmological perturbation
  theory}, \href{https://doi.org/10.1016/S0370-1573(02)00135-7}{\emph{Phys.
  Rep.} {\bfseries 367} (2002) 1}
  [\href{https://arxiv.org/abs/astro-ph/0112551}{{\ttfamily
  astro-ph/0112551}}].

\bibitem{martin_statistical_1973}
P.~C. Martin, E.~D. Siggia and H.~A. Rose, \emph{Statistical dynamics of
  classical systems}, \href{https://doi.org/10.1103/PhysRevA.8.423}{\emph{Phys.
  Rev. A} {\bfseries 8} (1973) 423}.

\bibitem{gozzi_hidden_1988}
E.~Gozzi, \emph{Hidden {BRS} invariance in classical mechanics},
  \href{https://doi.org/10.1016/0370-2693(88)90611-9}{\emph{Phys. Lett. B}
  {\bfseries 201} (1988) 525}.

\bibitem{gozzi_hidden_1989}
E.~Gozzi, M.~Reuter and W.~D. Thacker, \emph{Hidden {BRS} invariance in
  classical mechanics. {II}},
  \href{https://doi.org/10.1103/PhysRevD.40.3363}{\emph{Phys. Rev. D}
  {\bfseries 40} (1989) 3363}.

\bibitem{mazenko_fundamental_2010}
G.~F. Mazenko, \emph{Fundamental theory of statistical particle dynamics},
  \href{https://doi.org/10.1103/PhysRevE.81.061102}{\emph{Phys. Rev. E}
  {\bfseries 81} (2010) 061102}
  [\href{https://arxiv.org/abs/0905.4904}{{\ttfamily 0905.4904}}].

\bibitem{das_field_2012}
S.~P. Das and G.~F. Mazenko, \emph{Field theoretic formulation of kinetic
  theory: Basic development},
  \href{https://doi.org/10.1007/s10955-012-0610-y}{\emph{J. Stat. Phys.}
  {\bfseries 149} (2012) 643}
  [\href{https://arxiv.org/abs/1111.0571}{{\ttfamily 1111.0571}}].

\bibitem{bartelmann_microscopic_2016}
M.~Bartelmann, F.~Fabis, D.~Berg, E.~Kozlikin, R.~Lilow and C.~Viermann,
  \emph{A microscopic, non-equilibrium, statistical field theory for cosmic
  structure formation},
  \href{https://doi.org/10.1088/1367-2630/18/4/043020}{\emph{New J. Phys.}
  {\bfseries 18} (2016) 043020}
  [\href{https://arxiv.org/abs/1411.0806}{{\ttfamily 1411.0806}}].

\bibitem{bartelmann_kinetic_2017}
M.~Bartelmann, F.~Fabis, E.~Kozlikin, R.~Lilow, J.~Dombrowski and
  J.~Mildenberger, \emph{Kinetic field theory: effects of momentum correlations
  on the cosmic density-fluctuation power spectrum},
  \href{https://doi.org/10.1088/1367-2630/aa7e6f}{\emph{New J. Phys.}
  {\bfseries 19} (2017) 083001}
  [\href{https://arxiv.org/abs/1611.09503}{{\ttfamily 1611.09503}}].

\bibitem{fabis_kinetic_2018}
F.~Fabis, E.~Kozlikin, R.~Lilow and M.~Bartelmann, \emph{Kinetic field theory:
  exact free evolution of gaussian phase-space correlations},
  \href{https://doi.org/10.1088/1742-5468/aab850}{\emph{J. Stat. Mech.} (2018)
  043214} [\href{https://arxiv.org/abs/1710.01611}{{\ttfamily 1710.01611}}].

\bibitem{berges_non-perturbative_2002}
J.~Berges, N.~Tetradis and C.~Wetterich, \emph{Non-perturbative renormalization
  flow in quantum field theory and statistical physics},
  \href{https://doi.org/10.1016/S0370-1573(01)00098-9}{\emph{Phys. Rep.}
  {\bfseries 363} (2002) 223}
  [\href{https://arxiv.org/abs/hep-ph/0005122}{{\ttfamily hep-ph/0005122}}].

\bibitem{matarrese_resumming_2007}
S.~Matarrese and M.~Pietroni, \emph{Resumming cosmic perturbations},
  \href{https://doi.org/10.1088/1475-7516/2007/06/026}{\emph{J. Cosmol.
  Astropart. Phys.} {\bfseries 06} (2007) 026}
  [\href{https://arxiv.org/abs/astro-ph/0703563}{{\ttfamily
  astro-ph/0703563}}].

\bibitem{floerchinger_renormalization-group_2017}
S.~Floerchinger, M.~Garny, N.~Tetradis and U.~A. Wiedemann,
  \emph{Renormalization-group flow of the effective action of cosmological
  large-scale structures},
  \href{https://doi.org/10.1088/1475-7516/2017/01/048}{\emph{J. Cosmol.
  Astropart. Phys.} {\bfseries 01} (2017) 048}
  [\href{https://arxiv.org/abs/1607.03453}{{\ttfamily 1607.03453}}].

\bibitem{crocce_renormalized_2006}
M.~Crocce and R.~Scoccimarro, \emph{Renormalized cosmological perturbation
  theory}, \href{https://doi.org/10.1103/PhysRevD.73.063519}{\emph{Phys. Rev.
  D} {\bfseries 73} (2006) 063519}
  [\href{https://arxiv.org/abs/astro-ph/0509418}{{\ttfamily
  astro-ph/0509418}}].

\bibitem{bartelmann_trajectories_2015}
M.~Bartelmann, \emph{Trajectories of point particles in cosmology and the
  zel'dovich approximation},
  \href{https://doi.org/10.1103/PhysRevD.91.083524}{\emph{Phys. Rev. D}
  {\bfseries 91} (2015) 083524}
  [\href{https://arxiv.org/abs/1411.0805}{{\ttfamily 1411.0805}}].

\bibitem{mayer_molecular_1941}
J.~E. Mayer and E.~Montroll, \emph{Molecular distribution},
  \href{https://doi.org/10.1063/1.1750822}{\emph{J. Chem. Phys.} {\bfseries 9}
  (1941) 2}.

\bibitem{pietroni_coarse-grained_2012}
M.~Pietroni, G.~Mangano, N.~Saviano and M.~Viel, \emph{Coarse-grained
  cosmological perturbation theory},
  \href{https://doi.org/10.1088/1475-7516/2012/01/019}{\emph{J. Cosmol.
  Astropart. Phys.} {\bfseries 01} (2012) 019}
  [\href{https://arxiv.org/abs/1108.5203}{{\ttfamily 1108.5203}}].

\bibitem{park_cosmic_2000}
C.~Park, \emph{Cosmic momentum field and mass fluctuation power spectrum},
  \href{https://doi.org/10.1111/j.1365-8711.2000.03886.x}{\emph{Mon. Not. R.
  Astron. Soc.} {\bfseries 319} (2000) 573}
  [\href{https://arxiv.org/abs/astro-ph/0012066}{{\ttfamily
  astro-ph/0012066}}].

\bibitem{bardeen_statistics_1986}
J.~M. Bardeen, J.~R. Bond, N.~Kaiser and A.~S. Szalay, \emph{The statistics of
  peaks of gaussian random fields},
  \href{https://doi.org/10.1086/164143}{\emph{Astrophs. J.} {\bfseries 304}
  (1986) 15}.

\bibitem{crocce_memory_2006}
M.~Crocce and R.~Scoccimarro, \emph{Memory of initial conditions in
  gravitational clustering},
  \href{https://doi.org/10.1103/PhysRevD.73.063520}{\emph{Phys. Rev. D}
  {\bfseries 73} (2006) 063520}
  [\href{https://arxiv.org/abs/astro-ph/0509419}{{\ttfamily
  astro-ph/0509419}}].

\bibitem{ellis_tikz-feynman:_2017}
J.~P. Ellis, \emph{{TikZ}-feynman: Feynman diagrams with {TikZ}},
  \href{https://doi.org/10.1016/j.cpc.2016.08.019}{\emph{Comput. Phys. Commun.}
  {\bfseries 210} (2017) 103}
  [\href{https://arxiv.org/abs/1601.05437}{{\ttfamily 1601.05437}}].

\bibitem{valageas_new_2004}
P.~Valageas, \emph{A new approach to gravitational clustering: A path-integral
  formalism and large-n expansions},
  \href{https://doi.org/10.1051/0004-6361:20040125}{\emph{Astron. Astrophys.}
  {\bfseries 421} (2004) 23}
  [\href{https://arxiv.org/abs/astro-ph/0307008}{{\ttfamily
  astro-ph/0307008}}].

\bibitem{peebles_large-scale_1980}
P.~J.~E. Peebles, \emph{The large-scale structure of the universe}, Princeton
  series in physics. Princeton University Press, 1980.

\end{thebibliography}\endgroup

\end{document}